\documentclass[10pt]{iopart}
\usepackage{setstack,graphicx,color,iopams,bm,dirtytalk,float,algpseudocode,bbm}
\usepackage[mathscr]{euscript}

\newcommand{\black}{\color{black}}

\newcommand{\bA}{\mbox{\protect\boldmath $A$}}
\newcommand{\bC}{\mbox{\protect\boldmath $C$}}
\newcommand{\bD}{\mbox{\protect\boldmath $D$}}
\newcommand{\bx}{\mbox{\protect\boldmath $x$}}

\newcommand{\bz}{\mbox{\protect\boldmath $z$}}
\newcommand{\data}{\mathscr{D}}
\newcommand{\bra}{\langle}
\newcommand{\ket}{\rangle}
\newcommand{\pprime}{{\prime\prime}}
\newcommand{\rmD}{{\rm D}}

\newcommand{\order}{{\mathcal O}}
\newcommand{\vsp}{\vspace*{3mm}}

\begin{document}

\title[Overfitting in regression models for time to event data with censoring]{Replica analysis of overfitting in regression models for time to event data: the impact of  censoring} 

\author{E Massa$^{\dag}$, A Mozeika$^{\ast}$ and ACC Coolen$^{\dag\S}$}
\address{
$\dag$  Theoretical Biophysics, DCN Donders Institute, Faculty of Science,\\
~~Radboud University, 6525 AJ Nijmegen, The Netherlands\\[1mm]
$\ast$ Translational Oncology and Urology Research, King's College London, \\~~London SE1 9RT, United Kingdom \\[1mm]
$\S$ Saddle Point Science Europe, Mercator Science Park,  
\\~~6525 EC Nijmegen, The Netherlands}

\ead{emanuele.massa@donders.ru.nl, alexander.mozeika@kcl.ac.uk, a.coolen@science.ru.nl}

\begin{abstract}
We use statistical mechanics techniques, viz. the replica method, to model the effect of censoring on overfitting in Cox's proportional hazards model, the dominant regression method for time-to-event data. In the overfitting regime, Maximum Likelihood parameter estimators are known to be biased already for small values of  the ratio of the number of  covariates over the number of samples. 
The inclusion of censoring was avoided in previous overfitting analyses for mathematical convenience,  but is vital  to make any theory applicable to real-world medical data, where censoring is ubiquitous. Upon constructing efficient algorithms for solving the new (and more complex) RS equations and comparing the solutions with numerical simulation data, we find excellent agreement, even for large censoring rates.  
We then address the practical problem of using the theory to correct the biased ML estimators {without} knowledge of the data-generating distribution. This is  achieved via a novel numerical algorithm that self-consistently approximates all relevant parameters of the data generating distribution while simultaneously solving the RS equations.  We investigate numerically the statistics of the corrected estimators, and show that the proposed new algorithm indeed succeeds in removing the bias of the ML estimators, for both the association parameters and for the cumulative hazard.  

 \end{abstract}

\section{Introduction}
\label{section:Intro}

Inference relies on the foundations provided  by classical statistical theory, which was developed for use in settings where the number $p$ of covariates is small compared to the number $n$ of observations \cite{Asymptotic,kalbfleisch}. The standard Maximum Likelihood (ML) method for estimating model parameters indeed fails in the high-dimensional regime where $p=\mathcal{O}(n)$ \cite{el_karoui2013,donoho,coolen_2017,HD_logit,cgmt,loureiro}, due to overfitting. Overfitting is  the phenomenon that data noise is misinterpreted as signal, leading to  biased parameter estimators with large sample to sample fluctuations. An overfitting model will predict  outcomes for training examples well, but will fail in predicting outcomes for new data. 
Early strategies to mitigate overfitting include leaving out covariates (with the risk of overlooking relevant predictive information), penalization (equivalent to adding a parameter prior in Bayesian language), and shrinking parameter estimates after model fitting. The penalization weight or shrinking factor are typically estimated via bootstrapping or cross-validation, which forces one to sacrifice some of the training data; this makes the overfitting even worse. Moreover, penalization and shrinking may at best repair the overfitting-induced bias in the length of the parameter vector, but not the bias in its direction (which will appear as soon as the covariates are correlated \cite{GLM}).  

In order to use existing regression models also in the overfitting regime, it is vital to correctly quantify and undo the effects of overfitting, both for inference and for prediction purposes.
As a consequence, in recent years we have seen increased research efforts aimed at extending the classical inference theory to the so-called proportional asymptotic regime, characterized by taking the limit $n\to\infty$ while keeping the ratio  $\zeta=p/n$ fixed. Several mathematical methods from the domains of the statistical physics of disordered system \cite{coolen_2017,loureiro,GLM,massa}, computer science \cite{cgmt,cgmt_cox}  and statistics \cite{el_karoui2013,donoho,HD_logit}, have by now been applied successfully in this latter regime to model the statistics of ML and Maximum A Posteriori Probability (MAP) estimators. 

The condition $p\ll n$  for ML/MAP inference to be used safely is especially problematic in post-genome medicine:  we can now routinely measure very many variables per patient and want to use these to develop  personalized therapies, but overfitting prevents us from doing so. For time-to-event data,  the most common type in medicine, the main  regression tools are variations on the model of Cox \cite{Cox}. Overfitting in this model was analysed in \cite{coolen_2017,sheikh} as a statistical mechanics problem, via the replica method \cite{virasoro}. A later study \cite{cgmt_cox} used  the Convex Gaussian Min-max theorem \cite{cgmt}, with the logarithm of Cox's partial likelihood as utility function, to study overfitting in the association parameters.  The latter route avoids the use of replicas, but unlike 
 \cite{coolen_2017,sheikh}, cannot model overfitting  in the base hazard rate. 
 
All  studies of overfitting in Cox models have so far assumed for simplicity that the data were not subject to censoring. Censoring means that samples are lost to observation prior to events occurring. It  is ubiquitous in medicine, since medical studies are always of finite duration and also since patients often fail to return to hospital appointments for unknown reasons. 
Before the modern overfitting analysis methods for the proportional asymptotic regime, and their corresponding overfitting bias decontamination formulae, can be applied  in the real world, it is hence vital that the theory of \cite{coolen_2017,sheikh} is extended to include censoring.  That is the first aim of this paper. 

We generalise the theory of \cite{coolen_2017,sheikh} by allowing for arbitrary types of non-informative censoring, carry out a replica analysis in the regime $p,n\to \infty$ with fixed $\zeta=p/N$, and derive the extended RS equations. Methodologically we also improve upon  the approach in \cite{coolen_2017,sheikh} by: (i) using an more compact form of the replica approach, (ii)  avoiding the variational approximation for the base hazard rate, and (iii) constructing novel and more powerful numerical algorithms for solving and inverting the RS equations, including in cases where one does not know anything about the data generating distribution, resulting in precise algorithms for decontaminating estimators of association parameters and the integrated base hazard for overfitting-induced bias.    

This paper is organized as follows.
In Section \ref{section:setting} we define our notation and describe briefly the Cox model.
We explain the results obtained via the replica method (whose derivation is relegated to an Appendix) and the physical interpretation of the RS order parameters in Section \ref{section:replica_results}, and show in Section \ref{section:numerics} how this interpretation inspires an efficient numerical method for solving the RS  equations. In Section \ref{section:check_simulations} we test the predictions of the theory against numerical simulations, and find perfect agreement. We construct  in Section \ref{section:correction} a new algorithm for simultaneously inferring relevant characteristics of the data generating distribution and solving the RS equations, leading to a realistic and practical tool for correcting the biased ML estimates of association parameters and the nuisance parameters (i.e. the integrated base hazard) in the overfitting regime, even in the presence of censoring. We conclude in Section \ref{section:conclusion} with a discussion of our results and future research.

\section{Definitions}
\label{section:setting}

In time to event data, each observation ideally reports the time $T$ at which the subject experiences the event under investigation and the covariate vector $\bz = (z_{1},\dots,z_{p})\in \mathbb{R}^p$, i.e. the list of characteristics of the subject measured at time zero. All covariate vectors are assumed to have been drawn independently from some distribution $p(\bz)$. Since any non-zero average will drop out of our formulae, we can always take $p(\bz)$ to have average zero.   
However, in virtually all real-world time-to-event data sets observations are censored, i.e. for some subjects we have a missing or partial observation  of their event times. For instance, we might only know that an event occurred after or before a certain time point, or within a specific interval (called right, left and interval censoring, respectively).
In what follows we focus on right censoring, the most common form of censoring  in medical applications. In practice this amounts to assuming that when collecting data we actually observe
\begin{equation}
    t = \min\{T,C\}.
\end{equation}
The random variable $C$ models the censoring time, drawn randomly from an a priori unknown  distribution $p_c(x)$, and we are given a binary variable that indicates whether the subject has at time $t$ experienced the event ($\Delta = 1$) or has been censored ($\Delta = 0$):\begin{equation}
    \Delta = \cases{
      1  & if \ $T<C$\\
      0  & otherwise}
\end{equation}
Note that we can always write $p_{c}(x) = \lambda_c(x)\rme^{-\Lambda_c(x)}$, where $\lambda_c(x)$ is the censoring rate and $\Lambda_c(x)=\int_0^x\!\rmd x^\prime~\lambda_c(x^\prime)$. The Cox semi-parametric model \cite{Cox} assumes that the time at which a subject experiences the event is distributed according to  
\begin{equation}
\label{ph}
     p_T(x|\bz) = \lambda_0(x)\rme^{\bbeta_0\cdot\bz- \Lambda_0(x)\exp(\bbeta_0\cdot\bz)},
\end{equation}
where $\bbeta_0 \in \mathbb{R}^p$, $\lambda_0(x)$ is the true (non-negative) hazard rate, and $\Lambda_0(x)=\int_0^x\!\rmd x^\prime~\lambda_0(x^\prime)$ is the true cumulative hazard. 
The censoring times $C$ and the event times $T$ are assumed to be statistically independent, i.e. censoring time is said to be non-informative, and the joint distribution of observed times $t\geq 0$ and event type indicators $\Delta\in\{0,1\}$, given covariates $\bz$,  can then be written as
\begin{equation}
\label{density_Cox}
    p(t,\Delta|\bbeta_0\!\cdot\! \bz)  =  \big(\lambda_0(t)\rme^{\bbeta_0\cdot\bz}\big)^{\Delta}\big(\lambda_c (t)\big)^{1-\Delta}\rme^{- \Lambda_0(t)\exp(\bbeta_0\cdot\bz)-\Lambda_c(t)} \ .
    \label{eq:data_generation}
\end{equation}
The parameters one seeks to infer are the true vector $\bbeta_0$ and the true function $\lambda_0(t)$. 
It follows from (\ref{density_Cox}) upon computing  the log-likelihood density for a data-set of i.i.d. observations $\data=\{(t_i,\Delta_i,\bz_i)\}_{i=1}^n$,  that ML inference will give the estimators
\begin{eqnarray}
(\hat{\bbeta},\hat{\lambda})_{n}&=& {\rm argmax}_{\bbeta,\lambda}~ L_n\big(\bbeta,\lambda|\data\big)
\label{eq:full_ML}
\\
\label{ll_ph}
   L_n\big(\bbeta,\lambda|\data\big)&=& \sum_{i=1}^n\Big\{ \Delta_i\big(\log \lambda(t_i) +\bbeta\cdot\bz_i\big) - \Lambda(t_i)\rme^{\bbeta\cdot \bz_i} \Big\}\ .
\end{eqnarray}
The problem (\ref{eq:full_ML}) can be reduced by first maximizing over the hazard rate.  Setting the functional derivative with respect to $\lambda$ to zero gives an equation that can be solved for $\lambda(t)$, leading to the so-called Breslow estimator \cite{Cox}:
\begin{equation}
\label{breslow}
     \hat{\lambda}_{n}(t|\bbeta) = \frac{1}{n} \sum_{k=1}^n  \frac{\Delta_k\delta(t-t_k)}{n^{-1}\sum_{j=1}^n \theta(t_j - t_k)\rme^{\bbeta\cdot\bz_j} } \ ,
\end{equation}
where $\theta(x)$ denotes the step-function ($\theta(x>0)=1$ and $\theta(x<0)=0$). Substituting (\ref{breslow}) into (\ref{ll_ph}) and disregarding terms that are independ of $\bbeta$ we obtain 
\begin{eqnarray}
\hat{\bbeta}_{n}&=&  {\rm argmax}_{\bbeta}~L_n(\bbeta|\data)
\label{eq:ML_beta}
\\
L_n(\bbeta|\data)&=& 
\label{cox_lpl}
    \sum_{i=1}^n \Delta_i\Big\{\bbeta\cdot \bz_i -\log\Big(\frac{1}{n} \sum_{j=1}^n \Theta(t_j - t_i)\rme^{\bbeta\cdot\bz_j}\Big) \Big\} \ .
\end{eqnarray}
The right-hand side of (\ref{cox_lpl}) is the logarithm of the famous 
Cox partial likelihood \cite{Cox}.

\section{Typical behaviour of the Maximum (Partial) Likelihood estimator}
\label{section:replica_results}

Computing the estimator $\hat{\bbeta}_{\rm ML}$  in (\ref{eq:ML_beta}) is equivalent to finding the ground state of a fictitious physical system with Hamiltonian $\mathcal{H}_n(\bbeta|\data)=- L_n(\bbeta|\data)$, in which 
the association parameters $\bbeta$ act as the degrees of freedom, and the data-set $\data $ plays the role of quenched disorder.
Statistical physics thus allows us to compute the average over the disorder (i.e. the data $\data$) of the log-partial likelihood, giving the typical behaviour of the ML estimator,  in the proportional asymptotic regime  as
\begin{eqnarray}
\hspace*{-20mm}
 \lim_{n,p\rightarrow \infty,~\zeta = p/n} \Big\bra \frac{1}{n}L_n(\hat{\bbeta}_{\rm ML}|\data) \Big\ket_{\data}  &=& - \ \lim_{n,p\rightarrow \infty\:\ \zeta = p/n}\frac{1}{n} \Big\bra {\rm min}_{\bbeta}~\mathcal{H}_n(\bbeta|\data)\Big\ket_{\data}  \nonumber \\
    &&\hspace*{-20mm} = \lim_{n,p\rightarrow \infty,~\zeta = p/n}  \lim_{\gamma\rightarrow \infty} \frac{1}{n\gamma }\Big\bra \log \int\!\rmd\bbeta~ \rme^{-\gamma \mathcal{H}_n(\bbeta|\mathbf{T},\bz)} \Big\ket_{\data}  \ . 
    \label{eq:intermediate1}
\end{eqnarray}
In fact, for reasons that will become clear, we insert a constant regularization factor (whose impact will be zero due to the limit $\gamma\to\infty$), and replace (\ref{eq:intermediate1}) by 
\begin{eqnarray}
\hspace*{-20mm}
 \lim_{n,p\rightarrow \infty,~\zeta = p/n} \Big\bra \frac{1}{n}L_n(\hat{\bbeta}_{\rm ML}|\data) \Big\ket_{\data} &=&
 \nonumber
 \\[-2mm]
 &&\hspace*{-25mm} \lim_{n,p\rightarrow \infty,~\zeta= p/n}  \lim_{\gamma\rightarrow \infty} \frac{1}{n\gamma }\Big\bra \log \int\!\rmd\bbeta~ \rme^{\frac{1}{2}p\log p-\gamma \mathcal{H}_n(\bbeta|\mathbf{T},\bz)} \Big\ket_{\data}  \ . 
    \label{eq:intermediate}
\end{eqnarray}
The standard procedure to carry out the disorder average in parametric regression models via the replica method (which goes back to \cite{gardner}) builds on the assumption that the log-likelihood is a sum of $n$ independent terms, each depending on a single observation $(t_i,\Delta_i,\bz_i)$. For the present Hamiltonian $\mathcal{H}_n(\bbeta|\data)$ this is not the case, so we need an intermediate step. 
We introduce the empirical distribution 
\begin{equation}
    P_n(t,\Delta,h|\bbeta,\data) = \frac{1}{n}\sum_{i=1}^n \delta(t - t_i)\delta_{\Delta,\Delta_i} \delta(h - \bbeta\cdot\bz_i) 
    \label{eq:Pdensity}
\end{equation}
and observe that we can write the Hamiltonian (i.e. minus the log-partial likelihood) as the following functional:
\begin{eqnarray}
 \hspace*{-18mm}    \mathcal{H}\Big[P_n(.|\bbeta,\data )\Big] &=& n \mathcal{E}\Big[P_n(.|\bbeta,\data )\Big],
     \label{eq:E1}  
     \\[0.5mm]
  \hspace*{-18mm}    \mathcal{E}\Big[P(.)\Big] &=&\!\int\!\rmd t\rmd h~P(t,1,h)\Bigg[
     \log \sum_{\Delta^\prime}
     \int\!\rmd h^\prime\rmd t^\prime~  \theta(t^\prime-t) \rme^{h^\prime} P(t^\prime\!,\Delta^\prime\!,h^\prime)  
- h \Bigg].\nonumber
\\
\hspace*{-18mm}&&
     \label{eq:E2}
\end{eqnarray}
We can now write (\ref{eq:intermediate}) in a form where the disorder average effectively is computed over the density of states associated with (\ref{eq:Pdensity}). Upon introducing suitable delta-functionals one then achieves factorisation of the disorder average over the samples $i$ in the data set $\data$. 
The full replica derivation is given in \ref{appendix:replica} for completeness, and gives upon making the so-called replica-symmetric (RS) ansatz:
\begin{eqnarray}
\label{extr}
\hspace*{-10mm}   \lim_{n,p\rightarrow \infty\:\ \zeta = p/n} \Big\bra \frac{1}{n}\mathcal{H}_n(\hat{\bbeta}_{\rm ML}|\data)\Big\ket_{\data} =
\underset{u,v,w}{{\rm extr}}~\mathcal{F}(u,v,w) 
= \mathcal{F}(u_{\star},v_{\star},w_{\star}) 
\end{eqnarray}
with 
\begin{equation}
\hspace*{-20mm}
\label{optimal_cost}
    \mathcal{F}(u_\star,v_\star,w_\star)=\int\!{\rm D}y{\rm D}z \sum_{\Delta}\int\!\rmd t~p(t,\Delta|Sy)\Big[ \Lambda(t)\rme^{\xi_{\star}(t,\Delta,y,z) } -\Delta\xi_{\star}(t,\Delta,y,z) \Big],~~~
\end{equation}
\begin{eqnarray}
\label{eq:model_gauss}
p(t,\Delta|Sy)&=&\big(\rme^{Sy}\lambda_0(t)\big)^\Delta\lambda_c^{1-\Delta} (t)\rme^{- \Lambda_0(t)\exp(S y)-\Lambda_c(t)},
\\[1mm]
    \label{xi}
    \xi(t,\Delta,y,z) &=&w y + v z + u^2 \Delta - W\big(u^2\rme^{u^2\Delta + wy + vz}\Lambda(t)\big),
     \label{rs4}
     \end{eqnarray}
     and
     \begin{eqnarray}
     \hspace*{-25mm}
     \Lambda(s) &=&\! \int\!{\rm D}y\! \sum_{\Delta\in\{0,1\}}\int\!\rmd t~p(t,\Delta|Sy)\Bigg[ \frac{ \Delta\theta(s-t)}{\int\!{\rm D}y^\prime{\rm D}z^\prime\sum_{\Delta^\prime}\int_t^\infty\!\rmd t^\prime~p(t^\prime,\Delta^\prime|Sy^\prime)\rme^{\xi_{\star}(t^\prime,\Delta^{\prime},y^\prime,z^\prime)}}\Bigg].\nonumber
     \\  \hspace*{-20mm}&&
\end{eqnarray}
Here we used the standard short-hand ${\rm D}y=(2\pi)^{-\frac{1}{2}}\rme^{-\frac{1}{2}y^2}\rmd y$, and 
$W(x)$ is the Lambert $W$-function \cite{lambert_function}, defined by the equation $W(x)\exp[W(x)] = x$ for all $x$.
The only condition on $p(\bz)$  needed in the derivation of the above RS equations is that for $p\to\infty$ the true and inferred linear predictors $\bbeta\!\cdot\!\bz$ acquire Gaussian statistics\footnote{This will be trivially true if the covariate distribution $p(\bz)$ is itself Gaussian, but also if the conditions for the central limit theorem to hold apply (i.e. if the components of $\bz$ not too strongly correlated, and the components of $\bbeta$ are not too dissimilar in their scaling with $n$; see \cite{Feller}).}.  
The extremum $(u_\star,v_\star,w_\star)$ in (\ref{extr}) is found to satisfy the so-called RS equations:
\begin{eqnarray}
\hspace*{-15mm}
\zeta v &=& \int\!{\rm D}y{\rm D}z ~z\sum_\Delta\!\int\!\rmd t~ p(t,\Delta|Sy)~W(u^2\Lambda(t)\rme^{\Delta u^2+vz+wy}),
\label{eq:RS_v1}
\\
\hspace*{-15mm}
0&=&\int\!{\rm D}y{\rm D}z~y\sum_\Delta\!\int\!\rmd t~ p(t,\Delta|Sy)~\Big[W(u^2\Lambda(t)\rme^{\Delta u^2+vz+wy})\!-\!\Delta u^2\Big],
\label{eq:RS_w1}
\\
\hspace*{-15mm}
 \zeta v^2&=&\int\!{\rm D}y{\rm D}z\sum_\Delta\!\int\!\rmd t~ p(t,\Delta|Sy)\Big[W(u^2\Lambda(t)\rme^{\Delta u^2+vz+wy})\!-\!\Delta u^2\Big]^2.
\label{eq:RS_u1}
\end{eqnarray}
These can be written in alternative forms, for instance by using identities such as  $W(u^2\Lambda(t)\rme^{\Delta u^2+vz+wy})=u^2\Lambda(t)\rme^{\xi(t,\Delta,y,z)}=
\Delta u^2\!+\!vz\!+\!wy-\xi(t,\Delta,y,z)$, $W^\prime(x)=W(x)/x[1\!+\!W(x)]$, and via integration by parts over $z$ in (\ref{eq:RS_v1}), giving
\begin{eqnarray}
    \label{rs1}
    \zeta v^2 &=&\int\!{\rm D}y{\rm D}z \sum_{\Delta}\int\!\rmd t~p(t,\Delta|y)\Big[\xi(t,\Delta,y,z)\!-\!vz\!-\!wy\Big]^2, \\
      \label{rs2}
    1-\zeta &=& \int\!{\rm D}y{\rm D}z \sum_{\Delta}\int\!\rmd t~p(t,\Delta|y)\Big[\frac{1}{1+u^2 \Lambda(t)\rme^{\xi(t,\Delta,y,z)} }\Big],\\
    \label{rs3}
    w &=&
    \int\!{\rm D}y{\rm D}z\sum_{\Delta}\int\!\rmd t~p(t,\Delta|y) ~y\xi(t,\Delta,y,z).
\end{eqnarray}
For the distribution ${\mathcal P}(t,\Delta,h)$ and the functions  $\Lambda(t)$ and ${\mathcal S}(t)$ we find in RS ansatz the following expressions
  (see \ref{appendix:replica}):
         \begin{eqnarray}
  \mathcal{P}(t,\Delta,h)  &=&      
         \int\!{\rm D}y{\rm D}z~p(t,\Delta|Sy)~\delta\big[h-\xi(t,\Delta,y,z)\big], 
\label{eq:RS_P_final}
         \\
\Lambda(t)&=& \int\!\rmd t^\prime\rmd h^\prime~\frac{\theta(t-t^\prime){\mathcal P}(t^\prime,1,h^\prime)}{
{\mathcal S}(t^\prime)},
\label{eq:RS_Lambda_final}
\\
{\mathcal S}(t)&=& \sum_{\Delta^\prime}\int\!\rmd t^\prime \rmd h^\prime~\theta(t^\prime-t)\rme^{h^\prime}{\mathcal P}(t^\prime,\Delta^\prime,h^\prime).
\label{eq:RS_S_final} 
\end{eqnarray}
We note that an alternative way to derive  equations (\ref{eq:RS_v1},\ref{eq:RS_w1},\ref{eq:RS_u1}) (and a corresponding equation for $\Lambda(t)$) would have been to make appropriate choices such as $y\to (t,\Delta)$ and $\theta\to\{\lambda_0(t),\lambda_C(t)\}$ in the RS formulae of \cite{GLM}. However, that alternative route would not have generated the powerful expression (\ref{eq:RS_P_final}) found with the present approach. By retracing its derivation and making simple adaptations, (\ref{eq:RS_P_final}) can be generalized to 
\begin{eqnarray}
  {\mathcal P}(t,\Delta,h_0,h)  &=&      
  \frac{\rme^{-\frac{1}{2}h_0^2/S^2}} {S\sqrt{2\pi}}
  p(t,\Delta|h_0) 
           \int\!{\rm D}z~\delta\big[h-\xi(t,\Delta,y,z)\big], 
           \end{eqnarray}
           where
           \begin{eqnarray}
    \hspace*{-15mm}  {\mathcal P}(t,\Delta,h_0,h)  &=&     \lim_{n\to\infty} \frac{1}{n}\sum_{i=1}^n \Big\bra\delta(t - t_i)\delta_{\Delta,\Delta_i}
          \delta(h_0\! -\! \bbeta_0\cdot\bz_i)  \delta(h\! -\! \bbeta\cdot\bz_i) \Big\ket_{\!\data}
    \label{eq:Pdensity_general}
\end{eqnarray}
(with the standard convention that $\zeta=p/n$ is kept fixed in the limits $n,p\to\infty$). 
This result shows very transparently the impact of overfitting on the inferred risk factors $\bbeta\cdot\bz_i$, relative to  their true values $\bbeta_0\cdot\bz_i$. \vsp

At this point is is helpful to write $w=S\kappa$. 
The interpretation of the values $(\kappa_{\star},v_{\star})$ as solved from the RS equations above can be inferred directly from the results in \cite{GLM}: 
\begin{eqnarray}
    \kappa_{\star} = \lim_{p,n \rightarrow \infty }\big\bra \hat{\kappa}_n\big\ket_{\data},&~~~~~~&  \hat{\kappa}_n=\frac{\bbeta_0\!\cdot\!\bA\hat{\bbeta}_n}{\bbeta_0\!\cdot\!\bA\bbeta_0}
    \label{interpretation1}
    \\
     v^2_{\star} = \lim_{p,n \rightarrow \infty }\big\bra \hat{v}^2_n\big\ket_{\data}, && \hat{v}_n^2=\hat{\bbeta}_n\!\cdot\!\bA\hat{\bbeta}_n\! -\! \hat{\kappa}_{n}^2 \bbeta_0\!\cdot\!\bA\bbeta_0 
    \label{interpretation2}
\end{eqnarray}
with $p/n = \zeta$ fixed as $p,n\to\infty$. 
The importance of $(\kappa_\star,v_\star)$  derives from the facts that (asymptotically) the asymptotic distribution of $\hat{\bbeta}_n$ depends only these two quantities \cite{massa}, and that the overfitting induced inference bias and noise can be expressed in terms of $\kappa_\star$ and $v_\star$, respectively. 
The function $\Lambda(t)$ has the following interpretation: 
\begin{equation}
   \Lambda(t) = \lim_{p,n \rightarrow \infty }\int_0^t\!\rmd t^\prime  \Big\bra\hat{\lambda}_{\rm ML}(t^\prime)\Big\ket_{\!\data} 
\end{equation}
with $p/n=\zeta$ fixed, and with the Breslow estimator as given in (\ref{breslow}).

\section{Numerical solution of RS equations}
\label{section:numerics}

For fully parametric GLM models in the overfitting regime the RS equations can always be solved numerically in a relatively straightforward iterative manner by evaluating the relevant integrals via Gaussian quadratures.
Here, for the Cox model with censoring,  we have the added complication of the functional order parameter $\Lambda(t)$, which suggests we take a different approach inspired by population dynamics algorithms and our interpretation (\ref{eq:RS_P_final},\ref{eq:RS_Lambda_final},\ref{eq:RS_S_final}). For any given values of the scalar order parameters $(u,v,w)$, and given values of $S = |\bA^{1/2}\bbeta_0|$, we can generate $m\gg 1$ samples from the distribution 
\begin{eqnarray}
{\mathcal P}(t,\Delta,y,z|u,v,w)&=&     (2\pi)^{-1}\rme^{-\frac{1}{2}(y^2+z^2)}p(t,\Delta|Sy),
\end{eqnarray}
resulting in $\{(t_\ell,\Delta_\ell,y_\ell,z_\ell)\}_{\ell=1}^m$. We then define for each $(u,v,w,\Lambda)$ the quantities $\xi_\ell(u,v,w,\Lambda)=\xi(t_\ell,\Delta_\ell,y_\ell,z_\ell|u,v,w,\Lambda)$,  computed via   (\ref{xi}), and the estimator $\tilde{\Lambda}_m(t)$:
\begin{eqnarray}
\tilde{\Lambda}_m(t|u,v,w,\Lambda)&=& \sum_{\ell=1}^m \frac{\theta(t-t_\ell)\Delta_\ell}
{\sum_{k=1}^m  \theta(t_k-t_\ell)\rme^{\xi_k(u,v,w,\Lambda)}}
\end{eqnarray}
For sufficiently large population size $m$, the iterative algorithm defined by the mapping $\Lambda(.)\to \Lambda^\prime(.)=\tilde{\Lambda}_{m}(.|u,v,w,\Lambda)$ will now by construction have as its fixed-point the solution of (\ref{eq:RS_P_final},\ref{eq:RS_Lambda_final},\ref{eq:RS_S_final}). 
Furthermore, it is relatively easy to invert numerically equation (\ref{rs2}) and write $u$ for fixed $(v,w,\Lambda)$ as a function of $\zeta$,  for instance via Newton's method, so that $\zeta$ can always be chosen as the independent parameter of our RS equations (as it is always known). We can then solve the following surrogate set of RS equations: 
\begin{eqnarray}
\label{monte_rs1}
        \zeta v_m^2 &=& \frac{1}{m}\sum_{\ell=1}^m \Big(\xi_\ell-v_m z_\ell-w_m y_\ell\big)^2,\\
\label{monte_rs2}
        1-\zeta= &=& \frac{1}{m}\sum_{\ell=1}^m \frac{1}{1+u_m^2 \tilde{\Lambda}_m(t_\ell)\rme^{\xi_\ell}  },  \\
\label{monte_rs3}
        w_m  &=& \frac{1}{m}\sum_{\ell=1}^m  y_\ell\xi_\ell, \\
\label{monte_rs4}
        \tilde{\Lambda}_m(t_i) &=& \sum_{\ell=1}^m\frac{ \theta(t_i-t_\ell)\Delta_\ell}{\sum_{j=1}^m \theta(t_j - t_\ell)\rme^{\xi_j}},
\end{eqnarray}
where now
\begin{equation}
    \xi_i = u^2\Delta_i + v_mz_i +w_my_i - W\big(u_m^2\tilde{\Lambda}_m(t_i) \rme^{u_m^2\Delta_i +v_mz_i+w_my_i}\big).
\end{equation}
(e.g. via damped fixed point iteration). When simulating the data, $\Lambda_0(.)$ is known and, empirically, we find that substituting equation (\ref{monte_rs3}) with its equivalent version, obtained by Gaussian integration by parts of (\ref{rs3}),
\begin{equation}
    \fl w_m = S \frac{1}{m}\sum_{\ell=1}^m  \Big(u_m^2\Delta_\ell - W(u_m^2\tilde{\Lambda}_m(t_\ell)\rme^{\Delta_\ell u_m^2+v_mz+w_my})\Big)\Big(\Delta_\ell - \Lambda_0(t_\ell) \rme^{S y_{\ell}}\Big)/\zeta
\end{equation}
leads to much faster convergence with smaller population size and hence the latter is adopted instead of (\ref{monte_rs3}).

\section{Simulations tests of RS theory}

\label{section:check_simulations}
\begin{figure}[t]
\vspace*{-2mm}
\hspace*{-5.5mm}
\begin{minipage}{.55\textwidth}
\includegraphics[width=73mm,height=73mm]{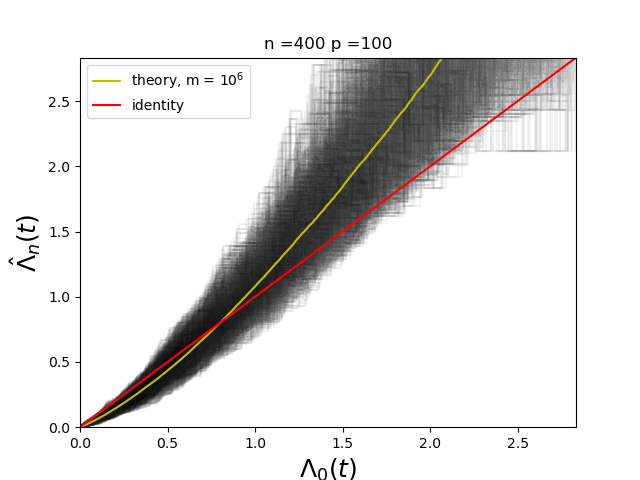}
\begin{center}\pt(a)\end{center}
\end{minipage}
\hspace*{-4mm}
\begin{minipage}{.55\textwidth}
\includegraphics[width=73mm,height=73mm]{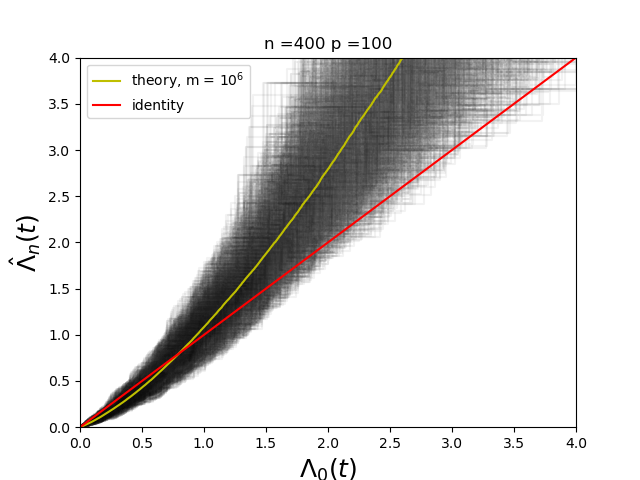}
\begin{center}\pt(b)\end{center}
\end{minipage}
\hspace*{-5.5mm}
\begin{minipage}{.55\textwidth}
\includegraphics[width=73mm,height=73mm]{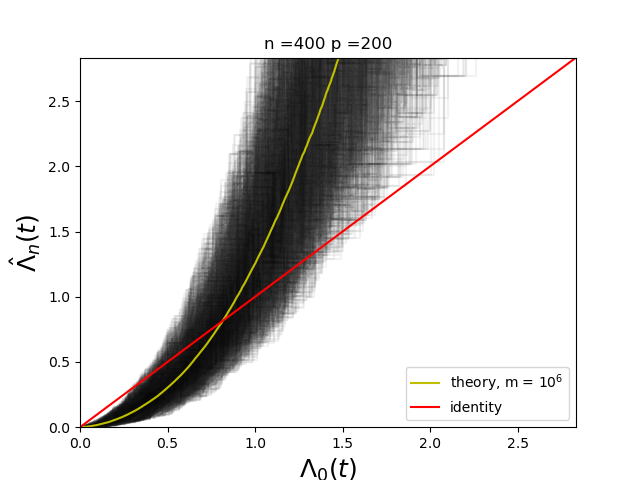}
\begin{center}\pt(c)\end{center}
\end{minipage}
\hspace*{-4mm}
\begin{minipage}{.55\textwidth}
\includegraphics[width=73mm,height=73mm]{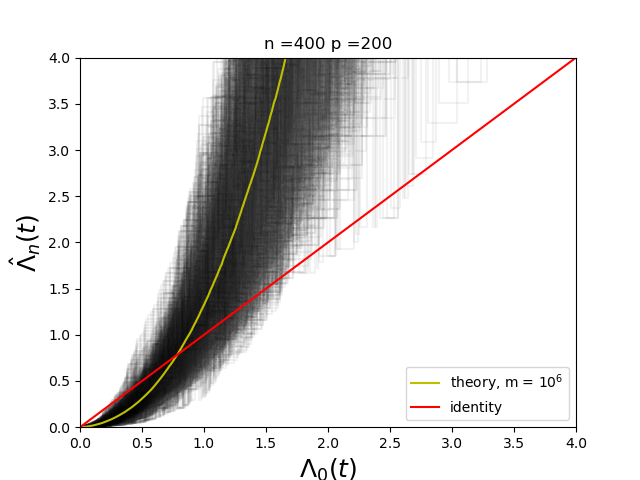}
\begin{center}\pt(d)\end{center}
\end{minipage}\vspace*{1mm} 
\caption{Comparison of theoretical predictions and simulation data, for Cox's survival analysis model with uniform censoring on the interval $[0,4]$, giving around 40\% censoring events, and data set size $n=400$. Top row: $p=100$ (so $\zeta=0.25$); bottom row: $p=200$ (so $\zeta=0.5$). Panels (a,c):  cumulative hazard  $\Lambda_0(t) = \log(1\!+\!t^2)$. Panels (b,d):  cumulative hazard $\Lambda_0(t) = \frac{1}{2}t^2$. In all panels we plot with black lines the Breslow estimator $\hat{\Lambda}_n(.)$ versus the true cumulative hazard $\Lambda_0(.)$, for  $500$ independent simulations with distinct data realizations. Yellow curves show the predictions of the RS theory (solved with populations of size $m=10^6$), and red curves indicate the diagonal (that would have been found for perfect regression). Further details are given in the main text.}
\label{fig:cox_cens_H}
\end{figure}

When some observations are censored, it is known that in the proportional asymptotic regime the ML estimator $\hat{\bbeta}_{\rm ML}$ does not always exists \cite{silvapulle}, and a recent paper \cite{cgmt_cox} established that, asymptotically and under the assumption of Gaussian covariates, 
 $\hat{\bbeta}_{n}$ undergoes a sharp phase transition at some critical value $\zeta_c$. Here we focus on the region $\zeta<\zeta_c$ where the ML estimator does exist. Since one obviously cannot solve the RS equations for all possible choices of the hazard function $\lambda_0(.)$, the censoring rate $\lambda_c(.)$  and the true association amplitude $S = |\bA^{1/2}\bbeta_0|$, we have in our simulations chosen a censoring distribution $p_c(t)$ that is uniform between $0$ and $t_{\rm max}$, reflecting the realistic scenario of a clinical trial of duration $t_{\rm max}$ where patients are recruited at constant rate for the trial duration. Different choices of $S$,   $t_{\rm max}$ and true hazard rate $\lambda_0(.)$ will lead to different expected fractions $\bra \Delta\ket$ of true (non-censored) events:
\begin{eqnarray}
\bra \Delta\ket&=& \int\!{\rm D}y\!\int_0^\infty \!\!\rmd t~p(t,1|Sy)\nonumber
=  -\int\!{\rm D}y\!\int_0^\infty \!\!\rmd t~\rme^{-\Lambda_c(t)}\frac{\rmd}{\rmd t}\rme^{-\Lambda_0(t)\exp(Sy)}
\nonumber
\\
&=& 1-\!\int_0^{t_{\rm max}}\!\!\frac{\rmd t}{t_{\rm max}}~\int\!{\rm D}y~\rme^{-\Lambda_0(t)\exp(Sy)}
\end{eqnarray}

\begin{figure}[t]
\hspace*{-5mm}
\begin{minipage}{.55\textwidth}
\includegraphics[width=\textwidth]{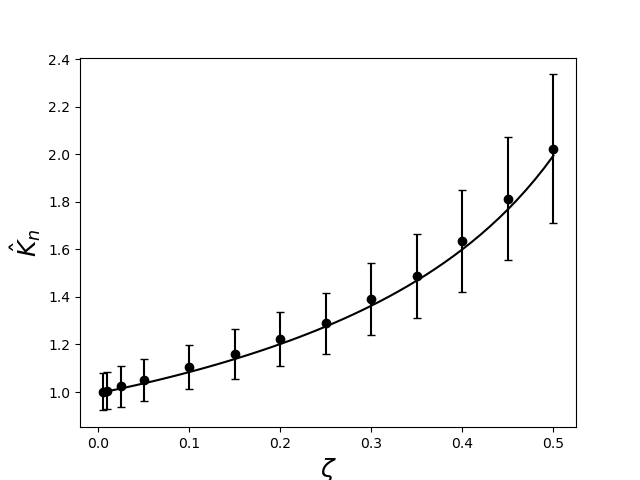}
\begin{center}\pt(a)\end{center}
\end{minipage}
\hspace*{-3mm}
\begin{minipage}{.55\textwidth}
\includegraphics[width=\textwidth]{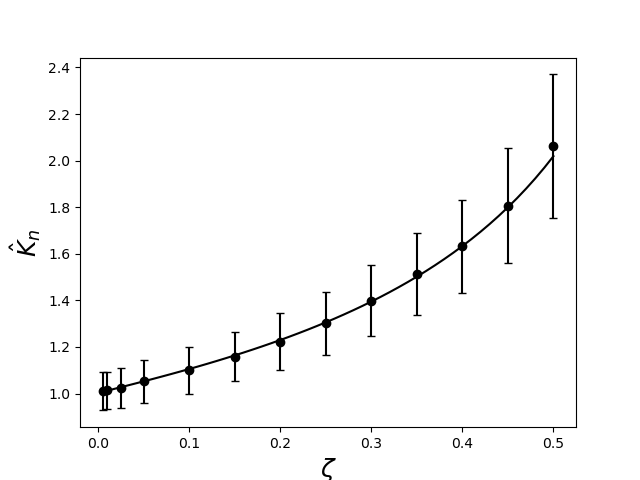}
\begin{center}\pt(b)\end{center}
\end{minipage}
\hspace*{-5mm}
\begin{minipage}{.55\textwidth}
\includegraphics[width=\textwidth]{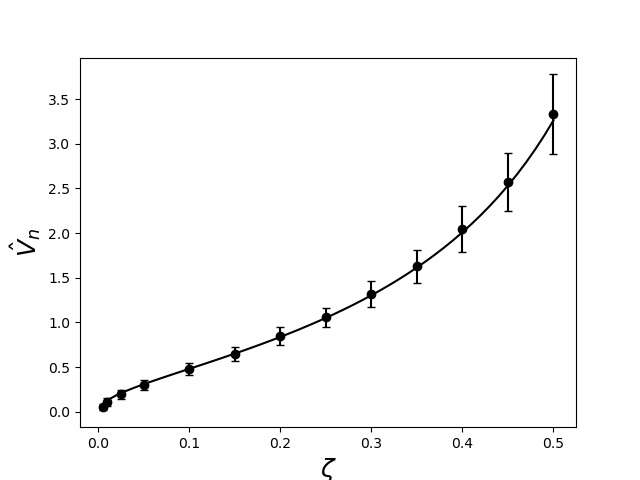}
\begin{center}\pt(c)\end{center}
\end{minipage}
\hspace*{-3mm}
\begin{minipage}{.55\textwidth}
\includegraphics[width=\textwidth]{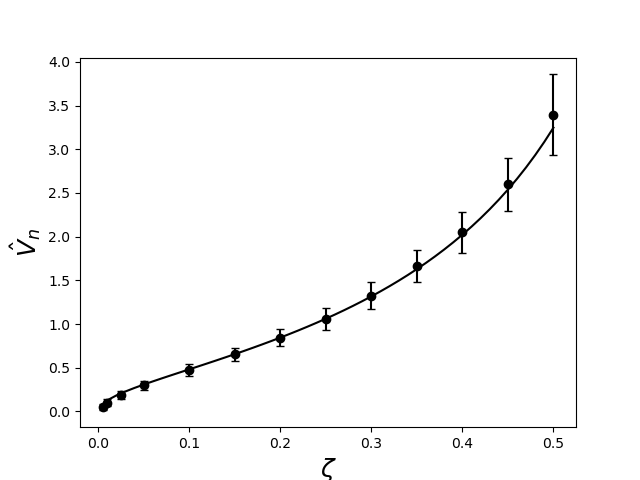}
\begin{center}\pt(d)\end{center}
\end{minipage}
\vspace*{1mm}
\caption{
Comparison of theoretical predictions and simulation data, for Cox's survival analysis model with uniform censoring on the interval $[0,4]$, giving around 40\% censoring events, and data set size $n=400$. Top row: overfitting-induced bias factor $\hat{\kappa}_n$ versus $\zeta$; bottom row: overfitting-induced noise amplitude $\hat{v}_n$ versus $\zeta$. Panels (a,c):  data for cumulative hazard  $\Lambda_0(t) = \log(1+t^2)$. Panels (b,d):  data for cumulative hazard $\Lambda_0(t) = \frac{1}{2}t^2$. In all panels we plot with markers and error bars the simulation results, averaged over $500$ independent simulations with distinct data realizations, and with solid lines the predictions $\kappa_\star$ and $v_\star$ of the RS theory (solved with populations of size $m=10^6$). } 
\label{fig:cox_cens_overlaps}
\end{figure}

To test the predictions of our RS equations we generated $500$ survival data-sets $\data$ of size $n=|\data|= 400$, with uniformly distributed censoring on the interval $[0,4]$, using cumulative hazards of the Log-logistic $\Lambda_0(t) = \log(1+t^2)$ and Weibull $\Lambda_0(t) = \frac{1}{2}t^2$ form. We used uncorrelated unit-variance Gaussian covariates, and true association vector $\bbeta_0 = \hat{\mathbf{e}}_1$ (i.e. the first unit vector), hence $S = 1$. In all cases $\bra \Delta\ket\approx 0.6$, so around 40\% events are censored, for both choices of $\Lambda_0(.)$. We then applied Cox's ML regression protocol to infer the associations and the cumulative hazard (via Breslow's formula), giving the estimators $\hat{\bbeta}_n$ and $\hat{\Lambda}_n(.)$. From $\hat{\bbeta}_n$ we subsequently computed the overfitting markers $(\hat{\kappa}_n,\hat{v}_n)$ in (\ref{interpretation1},\ref{interpretation2}).

In Figure \ref{fig:cox_cens_H} we plot the inferred cumulative hazards $\hat{\Lambda}_n(.)$ versus the true cumulative hazards $\Lambda_0(.)$, which can be done unambiguously since both are always non-decreasing functions of time. By the nature of the Breslow estimator (\ref{breslow}), these curves   take the form of `staircase' functions. We also show in the same panels the RS prediction of the relation between the two quantities, as red dashed lines. We conclude from Figure 1 that in all cases the solution of the RS equations correctly predict the typical values of  $\hat{\Lambda}_n(.)$. 
In Figure \ref{fig:cox_cens_overlaps} we compare the replica predictions $(\kappa_\star,v_\star)$ (solid lines) with the corresponding measurements $(\hat{\kappa}_n,\hat{v}_n)$ (markers with error bars), for different values of $\zeta=p/n\in[0,0.5]$.  Markers give the averages over the 500 data set realizations; error bars indicate the standard deviations. Again we observe excellent agreement between the RS theory and the simulations, in spite of the relatively small sample size $n = 400$. The latter feature is important for applications of the theory, since real data sets in survival analysis indeed often have sizes of that order.

\section{De-biasing protocols for overfitted ML estimators }
\label{section:correction}

Given the agreement between theory and simulations, we now explore the potential of using our theory as a systematic tool with which to decontaminate ML estimators and build {\em asymptotically unbiased} estimators $\tilde{\bbeta}_n$ and $\tilde{\Lambda}_n(.)$.
Two obstacles appear initially to hinder direct application of our RS equations for bias decontamination. First, our equations involve the amplitude $S$  (which is not directly accessible), Second, bias decontamination requires inverting the complex relation between the inferred and true integrated hazards. We have already constructed an algorithm for computing $\hat{\Lambda}_n(.)$ in terms of $\Lambda_0(.)$, but now we need to compute $\Lambda_0(.)$ given $\hat{\Lambda}_n(.)$. In this section we remove both obstacles, and show that the RS theory can indeed be used for creating accurate unbiased estimators 
in the overfitting regime. 

\subsection{Derivation of a practical algorithm for de-biasing}

Given the assumptions of our theory, we may write the marginal outcome probability $p(t,\Delta)=\int\!\rmd\bz~p(\bz)p(t,\Delta|\bbeta_0\!\cdot\!\bz)$ as \begin{eqnarray}
  p(t,\Delta) &=&\lambda_0^{\Delta}(t)\lambda_c^{1-\Delta}(t)\rme^{-\Lambda_c(t)}\int\!{\rm D}y~\rme^{\Delta Sy -  \Lambda_0(t)\exp(Sy)}.
\end{eqnarray}
Hence the log-marginal likelihood density for $n$ i.i.d. observations $\{(t_i,\Delta_i)\}$ reads
\begin{eqnarray}
\label{lml}
\hspace*{-22mm} 
   \ell_n(S,\lambda,\lambda_c) &=& \frac{1}{n} \sum_{i=1}^n\! \Big\{ \Delta_i\log \lambda(t_i)+ (1\!-\!\Delta_i)\log \lambda_c(t_i) - \Lambda_c(t_i) + \log \phi_{\Delta_i}\big(\Lambda(t_i),S \big) \Big\},
   \nonumber\\
   \hspace*{-20mm} &&
   \label{eq:marginal_LLD}
\end{eqnarray}
where we introduced the function
\begin{equation}
   \phi_{\Delta}(x,s) = \int\!{\rm D}y~\rme^{\Delta sy -x\exp(sy)}.
 \end{equation}
We next compute  the ML estimators for $(\lambda,\lambda_c)$ by maximization of  (\ref{eq:marginal_LLD}). Taking the functional derivative  with respect to $\lambda(.)$ gives, after standard manipulations, the following estimator for the  hazard rate: 
\begin{equation}
\label{lambda_frailty}
    \tilde{\lambda}(t|S) = \sum_{i=1}^n \frac{\Delta_i\delta(t-t_i)}{\sum_{j=1}^n \theta(t_j-t_i)\big[\phi_{\Delta_j+1}(\Lambda(t_j),S)/\phi_{\Delta_j}(\Lambda(t_j),S)\big]}.
\end{equation}
Upon integrating both sides over $t$, and replacing in the right-hand side the (as yet unknown) value of $\Lambda_(t_i)$ by the estimator   $\tilde{\Lambda}_n(t|S)=\int_0^t\!\rmd t^\prime~ \tilde{\lambda}_n(t^\prime|S)$ one then finds the following simple fixed-point equation for $\tilde{\Lambda}_n(.|S)$, that does not involve any parameter that could be susceptible to overfitting:
\begin{equation}
\label{integrated_lambda_frailty}
    \tilde{\Lambda}(t|S) = \sum_{i=1}^n \frac{\Delta_i\theta(t-t_i)}{\sum_{j=1}^n \theta(t_j-t_i)\big[\phi_{\Delta_j+1}(\tilde{\Lambda}(t_j),S)/\phi_{\Delta_j}(\tilde{\Lambda}(t_j),S)\big]}.
\end{equation}
The result of solving this fixed-point equation by (damped) iteration is remarkably accurate, as will be shown below.  
Repeating the same steps for the censoring rate $\lambda_c(.)$ gives the unbiased ML estimator
\begin{equation}
\label{Lambda_c}    \tilde{\Lambda}_c(t) = \sum_{i=1}^n \frac{(1\!-\!\Delta_i)\theta(t-t_i)}{\sum_{j=1}^n \theta(t_j-t_i)}.
\label{integrated_censoring_rate}
\end{equation}
\vsp

One could in principle attempt to determine $S$ in the same way: extremize (\ref{lml}) with respect to $S$, and solve the resulting equation simultaneously with (\ref{integrated_lambda_frailty}) for $\tilde{\Lambda}(.|S)$ and $S$. Unfortunately, the resulting equations admit the trivial solution $S= 0$, describing the trivial situation where the outcomes $(t,\Delta)$ are independent of the covariates. Thus we need an independent extra equation to determine the value of $S$, which must be added to and solved simultaneously with the RS order parameter equations and (\ref{integrated_lambda_frailty}). We first note that (\ref{interpretation1},\ref{interpretation2}) imply
\begin{eqnarray}
S^2=\frac{1}{\kappa_\star^2}\Big(\lim_{n,p\to\infty}\big\bra \hat{\bbeta}_n\!\cdot\!\bA\hat{\bbeta}_n\big\ket_{\data}-v_\star^2\Big)
\label{eq:get_S_sometimes}
\end{eqnarray}
with $p/n = \zeta$ fixed as $p,n\to\infty$. If the covariate correlation matrix $\bA$ is known, the above could serve as the desired extra equation. If the covariate correlation matrix is not known, as would generally be the case, we can use the following result, which exploits  the fact that in regression the $n$ quantities $\hat{\bbeta}_{\rm ML}\!\cdot\!\bz_i$ are always observable:
 \begin{eqnarray}
\lim_{n\to\infty} \frac{1}{n}\sum_{i=1}^n (\hat{\bbeta}_{\rm ML}\!\cdot\!\bz_i)^2&=& (\zeta\!-\!1)v_\star^2+w_\star^2.
\label{eq:get_S}
\end{eqnarray}
This identity is derived in \ref{app:get_S}. It uses explicitly the alternative form of the replica calculation followed in the present paper, from which followed equation (\ref{eq:RS_P_final}) (a result that could have been but was not derived in earlier papers). In contrast to (\ref{eq:get_S_sometimes}),  (\ref{eq:get_S}) can always be used as a additional  equation with which to determine $S$.

The final result is the following algorithm, that seeks to combine optimally the RS theory with the available data $\data$ and the biased ML estimators.   After measuring the left-hand side of (\ref{eq:get_S}), one solves numerically the four scalar equations (\ref{rs1},\ref{rs2},\ref{rs3},\ref{eq:get_S}) simultaneously for $(u,v,w,S)$, e.g. via (damped) fixed-point iteration, with the short-hands (\ref{rs4}) for the function $\xi(t,\Delta,y,z)$ and (\ref{eq:model_gauss}) for $p(t,\Delta|Sy)$. The  censoring rate $\lambda_c(t)$ in (\ref{eq:model_gauss}) is estimated by the time derivative of (\ref{integrated_censoring_rate}), and for each value of $S$, the base hazard $\lambda_0(t)$ by the time derive of the result of iterating (\ref{integrated_lambda_frailty}) to a fixed-point. The various Gaussian integrals can be done via Gauss-Legendre quadrature. The resulting new and unbiased estimator for the association parameters is according to   (\ref{interpretation1}) then given by $\tilde{\bbeta}=\hat{\bbeta}_{\rm ML}/\kappa_\star$, where $\kappa_\star=w_\star/S$.

\begin{figure}[t]
\vspace*{-2mm}
\hspace*{-5.5mm}\begin{minipage}{.55\textwidth}
\includegraphics[width=73mm,height=73mm]{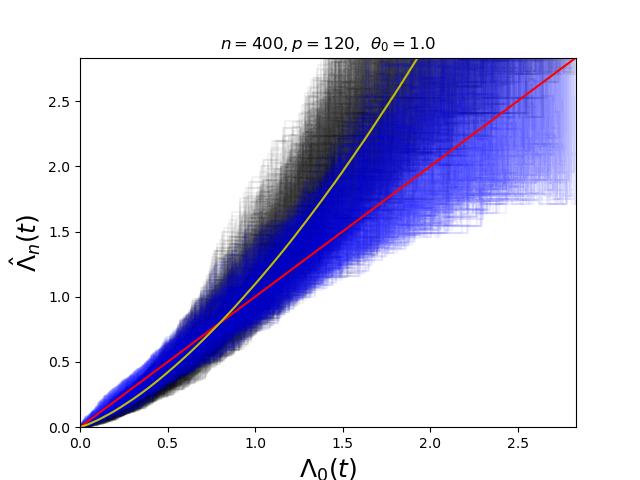}
\begin{center}\pt(a)\end{center}
\end{minipage}
\hspace*{-4mm}
\begin{minipage}{.5\textwidth}
\includegraphics[width=73mm,height=73mm]{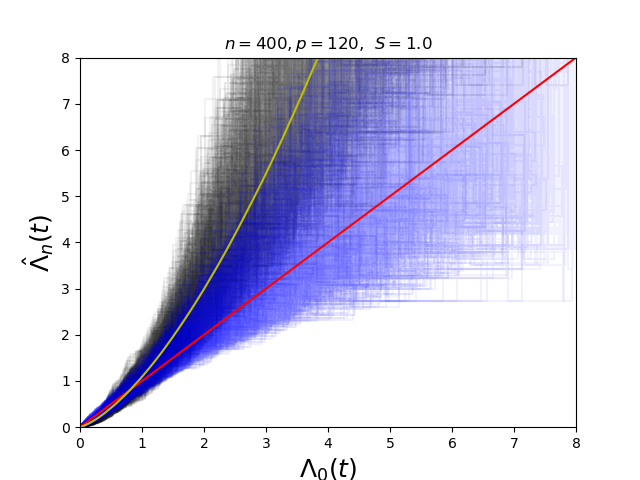}
\begin{center}\pt(b)\end{center}
\end{minipage}
\hspace*{-5.5mm}
\begin{minipage}{.55\textwidth}
\includegraphics[width=73mm,height=73mm]{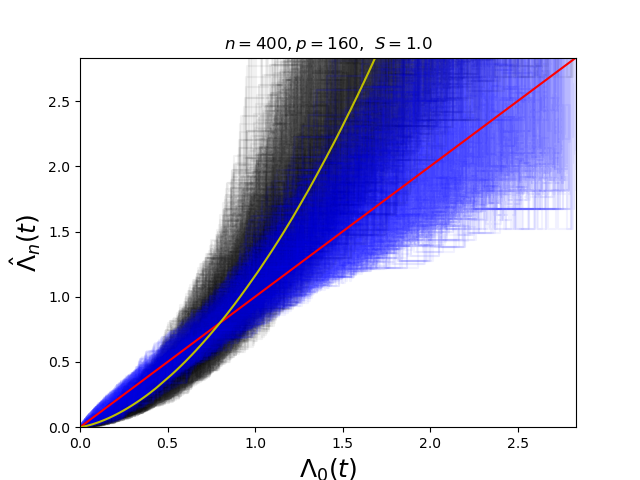}
\begin{center}\pt(c)\end{center}
\end{minipage}
\hspace*{-4mm}
\begin{minipage}{.5\textwidth}
\includegraphics[width=73mm,height=73mm]{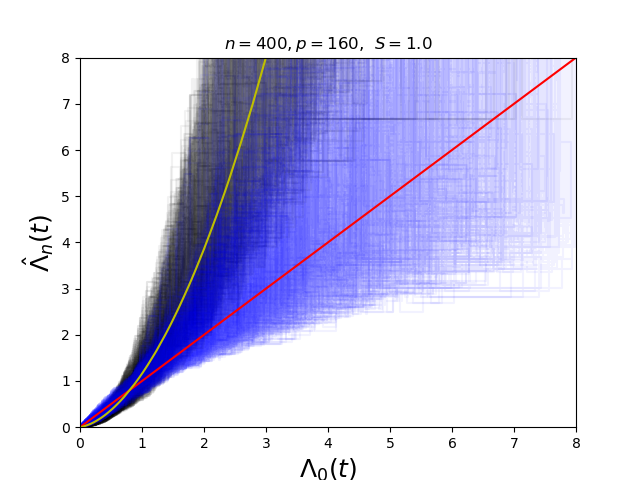}
\begin{center}\pt(d)\end{center}
\end{minipage}
\caption{
Comparison of uncorrected and corrected integrated hazard rate estimators derived from simulation data, for Cox's survival analysis model with uniform censoring on the interval $[0,4]$, giving around 40\% censoring events, and data set size $n=400$. In all cases $\bbeta_0 = \hat{\mathbf{e}}_1$, so $S=1$. 
Top row: $p=120$ (so $\zeta=0.3$); bottom row: $p=200$ (so $\zeta=0.4$). Panels (a,c):  cumulative hazard  $\Lambda_0(t) = \log(1\!+\!t^2)$. Panels (b,d):  cumulative hazard $\Lambda_0(t) = \frac{1}{2}t^2$. In all panels we plot with black lines the Breslow estimator $\hat{\Lambda}_{n}(.)$ versus the true cumulative hazard $\Lambda_0(.)$, for  $500$ independent simulations with distinct data realizations. Yellow curves show the predictions of the RS theory (solved with populations of size $m=10^6$), and red curves indicate the diagonal (that would have been found for perfect regression). 
We also show in blue the {\em de-biased estimator} $\tilde{\Lambda}(.)=\hat{\Lambda}_{n}(.)/\kappa_\star$ for the cumulative hazard versus the true cumulative hazard $\Lambda_0(.)$, for the same  $500$ simulations. 
}
\label{fig:cox_cens_corrected_H}
\end{figure}

\clearpage

\subsection{Simulation tests of the proposed new estimators}

\begin{figure}[t]
\begin{minipage}{.5\textwidth}
\includegraphics[width=\textwidth]{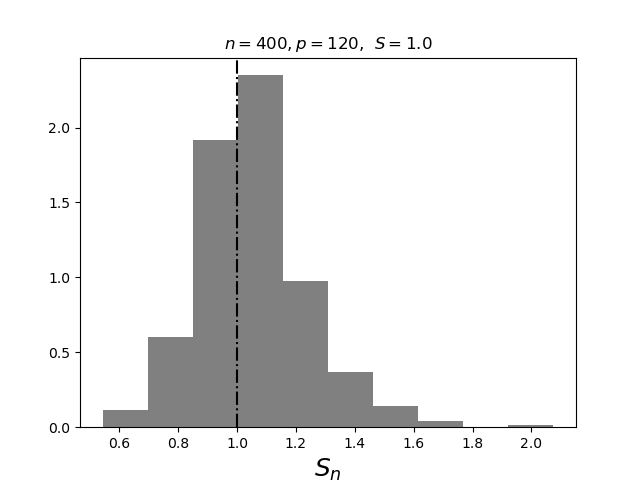}
\begin{center}\pt(a)\end{center}
\end{minipage}
\hfill
\begin{minipage}{.5\textwidth}
\includegraphics[width=\textwidth]{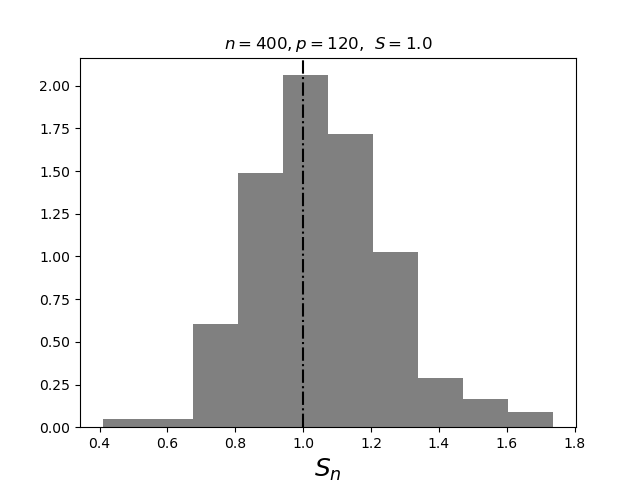}
\begin{center}\pt(b)\end{center}
\end{minipage}
\medskip
\begin{minipage}{.5\textwidth}
\includegraphics[width=\textwidth]{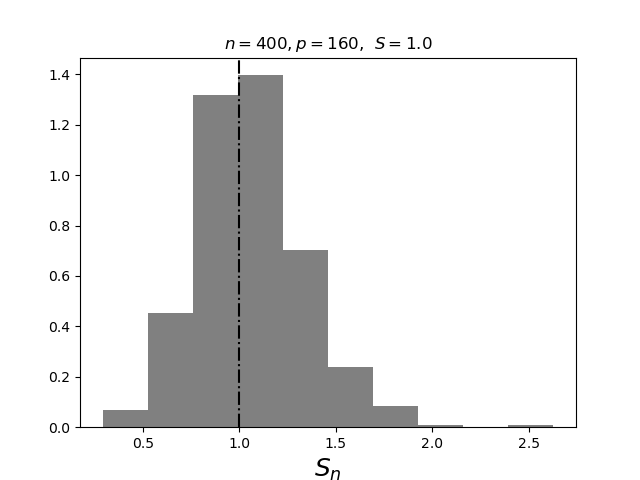}
\begin{center}\pt(c)\end{center}
\end{minipage}
\hfill
\begin{minipage}{.5\textwidth}
\includegraphics[width=\textwidth]{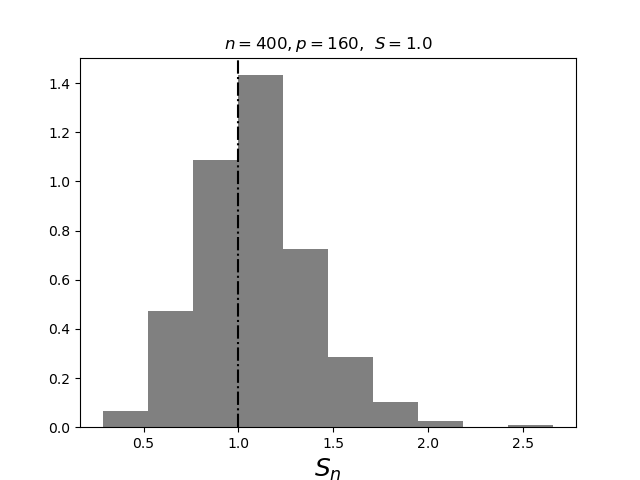}
\begin{center}\pt(d)\end{center}
\end{minipage}
\caption{Distributions of the values of  the effective signal strength $S= |\bA^{1/2}\bbeta_0|$ inferred by our de-biasing algorithm, for the same data as those used in Figure 3.  In all panels the maximum of the histogram corresponds to the true value $S=1$, in spite of the relatively modest data set size (around 240 non-censored events). 
} 

\label{fig:cox_cens_corrected_theta}
\end{figure}

\begin{figure}[t]
\begin{minipage}{.5\textwidth}
\includegraphics[width=\textwidth]{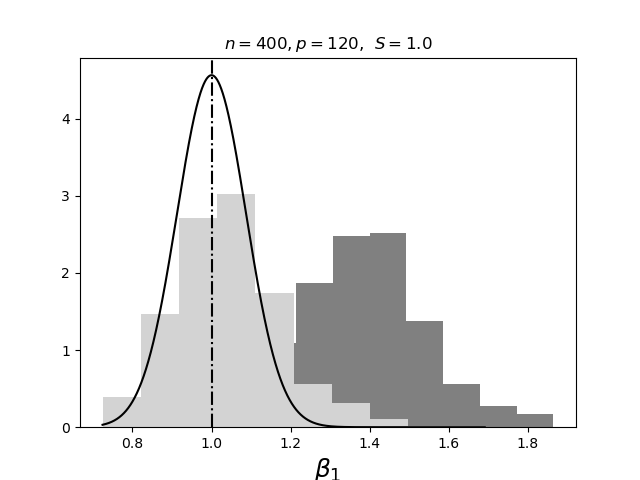}
\begin{center}\pt(a)\end{center}
\end{minipage}
\hfill
\begin{minipage}{.5\textwidth}
\includegraphics[width=\textwidth]{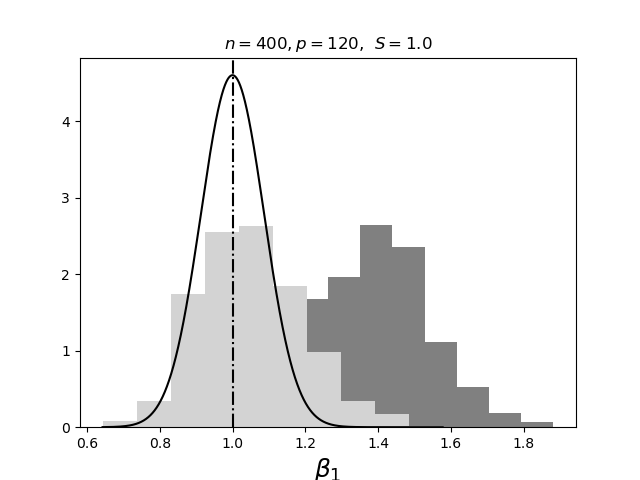}
\begin{center}\pt(b)\end{center}
\end{minipage}
\medskip
\begin{minipage}{.5\textwidth}
\includegraphics[width=\textwidth]{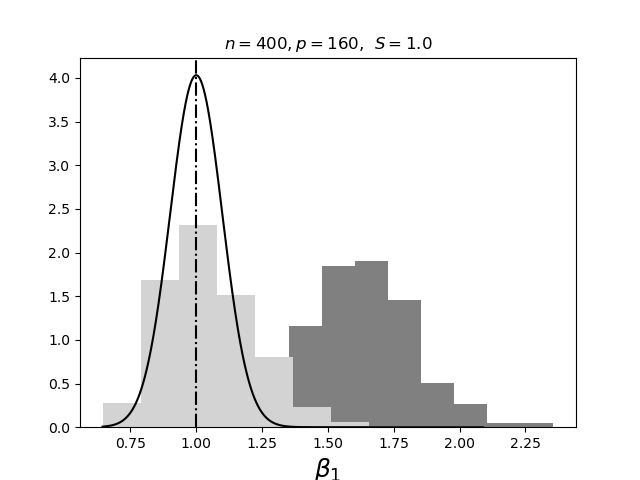}
\begin{center}\pt(c)\end{center}
\end{minipage}
\hfill
\begin{minipage}{.5\textwidth}
\includegraphics[width=\textwidth]{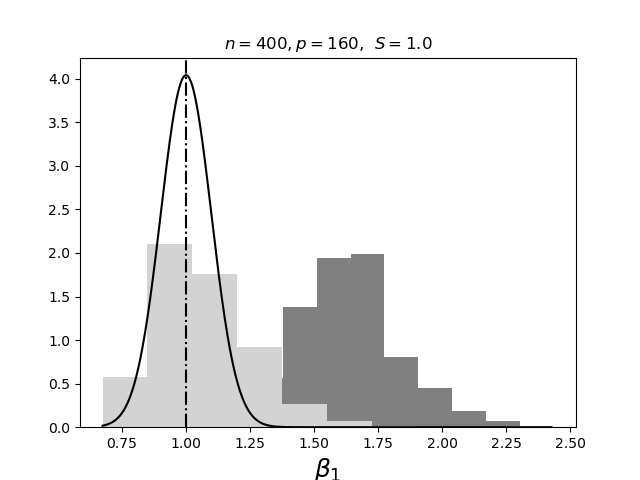}
\begin{center}\pt(d)\end{center}
\end{minipage}
\caption{ 
Distributions of the values of  the non-zero component $\hat{\bf e}_1\!\cdot\!\bbeta$ of the association vector, both for the standard ML Cox regression estimator (dark grey) and for the new estimator inferred by our de-biasing algorithm (light grey), for the same data as those used in Figure 3. We also show as a vertical dashed line the location of the true value $\hat{\bf e}_1\!\cdot\!\bbeta_0=1$, and as a solid curve the predicted Gaussian asymptotic distribution of the new estimator, with average 1 and variance  $v_{\star}^2/ \kappa^2_{\star}p$.  We observe that the new estimator indeed removed successfully the overfitting-induced bias of the ML one, while its distribution exhibits finite size effects. 
} 
\label{fig:cox_cens_corrected_beta1}
\end{figure}

\begin{figure}[t]
\begin{minipage}{.5\textwidth}
\includegraphics[width=\textwidth]{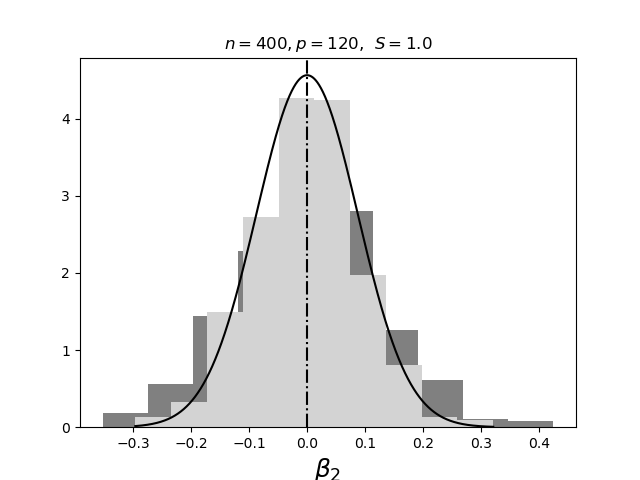}
\begin{center}\pt(a)\end{center}
\end{minipage}
\hfill
\begin{minipage}{.5\textwidth}
\includegraphics[width=\textwidth]{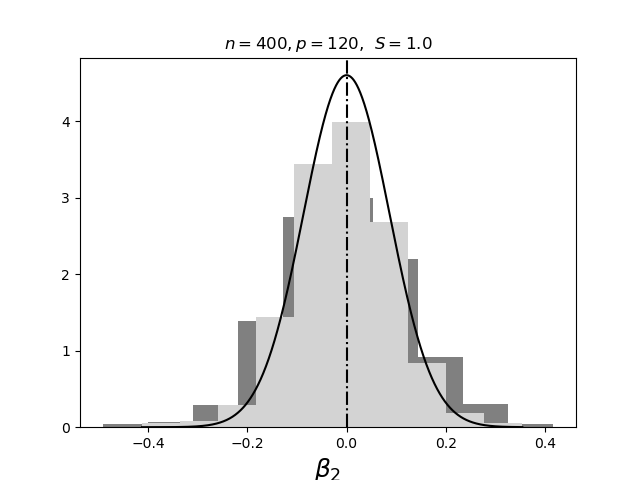}
\begin{center}\pt(b)\end{center}
\end{minipage}
\medskip
\begin{minipage}{.5\textwidth}
\includegraphics[width=\textwidth]{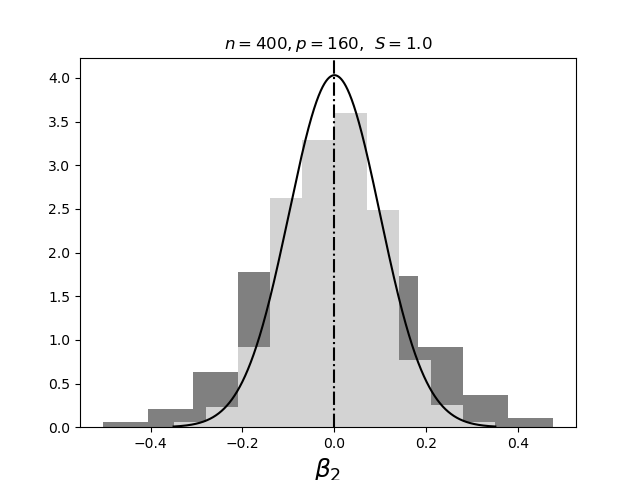}
\begin{center}\pt(c)\end{center}
\end{minipage}
\hfill
\begin{minipage}{.5\textwidth}
\includegraphics[width=\textwidth]{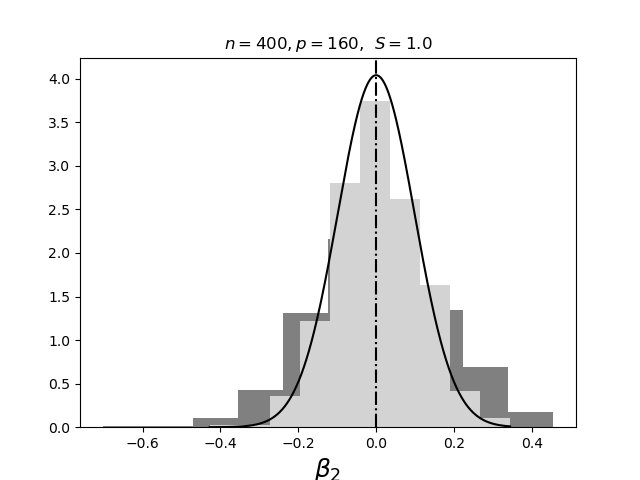}
\begin{center}\pt(d)\end{center}
\end{minipage}
\caption{ 
Distributions of the values of  the zero component $\hat{\bf e}_2\!\cdot\bbeta$ of the association vector, both for the standard ML Cox regression estimator (dark grey) and for the new estimator inferred by our de-biasing algorithm (light grey), for the same data as those used in Figure 3. We also show as a vertical dashed line the location of the true value $\hat{\bf e}_2\!\cdot\!\bbeta_0=0$, and as a solid curve the predicted Gaussian asymptotic distribution of the new estimator, here with zero average and variance  $v_{\star}^2/ \kappa^2_{\star}p$.   Here, as expected, already the ML estimator was unbiased, since the theory predicts the overfitting-induced bias to take the form of a multiplicative factor $\kappa_\star$ (which would here just multiply the number zero). 
} 
\label{fig:cox_cens_corrected_beta2}
\end{figure}

In Figure \ref{fig:cox_cens_corrected_H} we plot both the uncorrected and the de-biased estimator for the cumulative hazard, for simulations of survival data with $n=400$ and around $40\%$ of censoring events. These data show convincingly that the ML estimator $\hat{\Lambda}_{n}(.)$ (shown in black) is quite biased, with average values as predicted by the RS theory (dashed yellow line), but that the corrected estimator $\tilde{\Lambda}(.)$ (shown in blue) when plotted against the true cumulative hazard is  centered around the red dashed line (the diagonal, corresponding to unbiased estimation). This is remarkable,  given that our de-biasing algorithm does not require any knowledge of the data generating process, apart from assuming the form of a semi-parametric proportional hazard model (\ref{density_Cox}). As expected, the corrected estimator exhibits a larger variance than the ML estimator (reflecting the general and well-known bias-variance trade-off in inference). The increasing variance of both estimators at large times is simply a finite size effect: already at $t=2.5$ the survival probability is below $0.1$, so only few subjects are still alive and able to generate events that may contribute to hazard rate inference. 
Similarly we show in Figures \ref{fig:cox_cens_corrected_theta},  \ref{fig:cox_cens_corrected_beta1}, and \ref{fig:cox_cens_corrected_beta2} the differences between the ML estimators and the new de-biased estimators for the amplitude $S$ (the `signal strength' of the true associations) and the nonzero end zero components of the true association vector. 
 
 Our simulations support the conclusion that the new decontamination algorithm proposed above indeed leads to virtually unbiased estimators for $\Lambda_0(.)$, $S$ , and $\bbeta_0$, even for relatively small data sets  and in the presence of significant levels of censoring. 
This is a nontrivial result, since we have only required that there is no model mismatch, i.e. that the data were generated from  a Cox model, and that the distribution of the risk factors $\bbeta_0\cdot\bz_i$ will asymptotically be Gaussian. In particular, the algorithm does not require  a priori knowledge of any of the parameters of the data generating process.

\section{Conclusion}
\label{section:conclusion}

In this paper we have extended previous analytical studies on overfitting in high dimensional Cox regression for time-to-event data \cite{coolen_2017,GLM,sheikh}, by including  censoring. Censoring is a necessary ingredient for any survival analysis theory if its results are to be applied to  real medical data (the main application area of survival analysis). While still using the replica method as our main tool, in addition to the inclusion of censoring we have here also chosen a different route to carry out the analysis compared to \cite{coolen_2017,GLM,sheikh}, which enabled us to derive an explicit formula for the inferred distribution of risk factors. 
This latter formula, in turn, paved the way for the creation of new algorithms for applications of the replica-symmetric (RS) theory,  that previously required additional knowledge and/or additional approximations. All our new theoretical results and predictions are validated using numerical simulations. 

We consider the main deliverables of the present paper to be the following three. First, we now have an accurate theory with which to understand quantitatively the impact of censoring on overfitting phenomena in the Cox model, so that replica-based overfitting analysis can now be applied to realistic real-world problems in medicine (i.e. to data from clinical studies of finite as opposed to infinite duration).  Second, we are now able to solve directly and accurately the RS equations for the inferred cumulative hazard, instead of using the variational approximations proposed in  \cite{coolen_2017,GLM,sheikh}. 
Thirdly, we have been able to construct from the present theory a practical and accurate  algorithm with which to decontaminate inferences for overfitting-induced bias, that does not require knowledge of {\em any} of the parameters of the true data generating distribution. The latter algorithm combines in a very effective way the RS equations with those quantities that are measurable in ML regression, and infers the effective signal strength $S =|\bA^{1/2}\bbeta_0|$ as a by-product.
 
The results presented here thus have not only a theoretical but also a very practical value. The availability of tools for de-biasing the important estimators of the Cox model implies that one can now achieve significantly improved outcome predictions in realistic clinical scenarios, compared to what would have been achievable with ML regression alone, either by using more covariates for prediction (i.e. increasing $\zeta=p/n$ by increasing $p$), or by using fewer patients to do so (i.e. increasing $\zeta=p/n$ by decreasing $n$), or by allowing for increased levels of censoring without detrimental consequences. To the best of our knowledge, the presently proposed bias decontamination algorithm, that corrects both  association parameters and nuisance parameters (i.e. the cumulative hazard in the Cox model),  is the first of its kind. 

We envisage future work to involve application of the new estimator decontamination algorithm to real (partially censored) survival data, finding an expression or equation for the transition point $\zeta_c$ of the Cox model in the presence of censoring, and working out the RS theory upon including ridge penalization in the Cox formulae for survival analysis with censoring. We also envisage generalizing the new decontamination algorithm to more general GLM regression models, where the efficient computation of the signal strength $S$, of other possible parameters of the true data generating model,   and of the nuisance parameters, have in the past always required unwelcome approximations that we expect will now no longer be necessary.

\vspace*{3mm}\black

\noindent{\bf Acknowledgements}\\[1mm]
 AM  would like to thank  Mansoor Sheikh and Fabian Aguirre Lopez for constructive discussions. This work was partially supported by the Medical Research Council of the  United Kingdom (grant MR/L01257X/1).

\appendix

\section{Replica derivation}
\label{appendix:replica} 

We start from formulae (\ref{eq:Pdensity},\ref{eq:E1},\ref{eq:E2}) for the Hamiltonian, being minus the partial log-likelihood of the Cox model. 
The relevant  information on the typical estimators $(\hat{\bbeta},\hat{\lambda})_{\rm ML}$ follows from the asymptotic disorder-averaged free energy density in the $\gamma\to\infty$ limit, viz. from $f=\lim_{\gamma\to\infty}f(\gamma)$, where 
\begin{eqnarray}
    f(\gamma) &=& -\!\lim_{n\rightarrow \infty} \frac{1}{n\gamma} \Big\bra\! \log Z_n(\gamma,\data)\Big\ket_{\!\data},
    \\
    Z_n(\gamma,\data)&=& \int\!\rmd\bbeta~ \rme^{\frac{1}{2}p\log p-\gamma  \mathcal{H}[P_n(.|\bbeta,\data)]}.
\end{eqnarray}
To compute the average of the logarithm we use the replica method, giving
\begin{eqnarray}
   f(\gamma) &=& -\lim_{n\rightarrow \infty} \lim_{r\rightarrow 0}  \frac{1}{n\gamma r}\log \Big\bra [Z_n(\gamma,\data)]^r\Big\ket_{\data}.
  \label{eq:FED}
 \end{eqnarray}

\subsection{The replicated free energy density}

In order to compute  $\big\bra [Z_n(\gamma,\data)]^r\big\ket_{\data}$ 
we first write $Z_n(\gamma,\data)$ in a more convenient form.
We introduce the functional delta distribution
\begin{eqnarray} 
\hspace*{-20mm}
    \delta\Big[\mathcal{P}(.) \!-\! P_n(.|\bbeta,\data)\Big] &=& \int\!\mathcal{D}\hat{\mathcal{P}}~ \rme^{\rmi n\sum_{\Delta} \int\!\rmd t\rmd h~ \hat{\mathcal{P}}(t,\Delta,h)\big[\mathcal{P}(t,\Delta,h) - P_n(t,\Delta,h|\bbeta,\data)\big]  } 
    \end{eqnarray}
in which $\mathcal{D}\hat{\mathcal{P}}$ and $\mathcal{D}\mathcal{P}$ are normalized such that  $\int\!\mathcal{D}{\mathcal{P}}~ \delta\big[\mathcal{P}(.) \big] =1$. 
We can now write 
\begin{eqnarray}
   Z_n(\gamma,\mathcal{D})&=& \int\!\mathcal{D}\hat{\mathcal{P}} \mathcal{D}\mathcal{P}~\rme^{ n \big\{ \rmi\sum_{\Delta} \int\rmd t\rmd h~ \hat{\mathcal{P}}(t,\Delta,h)\mathcal{P}(t,\Delta,h) - \gamma \mathcal{E}[\mathcal{P} (.)] \big\} }\nonumber \\[-0.5mm]
    && \times\int\!\rmd\bbeta~ \rme^{\frac{1}{2}p\log p- \rmi \sum_{i=1}^n \hat{\mathcal{P}}(t_i, \Delta_i,\bbeta\cdot\bz_i) } .
\end{eqnarray}
For integer $r$ we can next do the average over the data $\data$: 
\begin{eqnarray}
\hspace*{-23mm}  \big\bra[Z_n(\gamma,\data)]^r\big\ket_{\!\data}&=&
\! \int\!\Big[\prod_{\alpha=1}^r\! \mathcal{D}\hat{\mathcal{P}}_\alpha \mathcal{D}\mathcal{P}_\alpha\Big]
 ~\rme^{ n \sum_{\alpha=1}^r\big\{ \rmi\sum_{\Delta} \int\rmd t\rmd h~ \hat{\mathcal{P}}_\alpha(t,\Delta,t)\mathcal{P}_\alpha(t,\Delta,t) - \gamma \mathcal{E}[\mathcal{P}_\alpha (.)] \big\} } 
 \nonumber
 \\
 &&\hspace*{-15mm} \times\!
 \int\!\Big[\prod_{\alpha=1}^r \rmd\bbeta_\alpha\Big]\rme^{\frac{1}{2}rp\log p}\Big[\sum_{\Delta}\!\int\!\rmd\bz\rmd t~p(t,\Delta|\bbeta_0\!\cdot\!\bz)p(\bz) \rme^{- \rmi \sum_{\alpha=1}^r \hat{\mathcal{P}}_\alpha(t, \Delta,\bbeta_\alpha\cdot\bz) }\Big]^n 
 \nonumber
 \\
 &&\hspace*{-7mm}= \int\!\Big[\prod_{\alpha=1}^r\! \mathcal{D}\hat{\mathcal{P}}_\alpha \mathcal{D}\mathcal{P}_\alpha\Big]
 ~\rme^{ n \sum_{\alpha=1}^r\big\{ \rmi\sum_{\Delta} \int\rmd t\rmd h~ \hat{\mathcal{P}}_\alpha(t,\Delta,h)\mathcal{P}_\alpha(t,\Delta,h) - \gamma \mathcal{E}[\mathcal{P}_\alpha (.)] \big\} } 
 \nonumber
 \\
 &&\hspace*{-14mm} \times\!
 \int\!\Big[\prod_{\alpha=1}^r \rmd\bbeta_\alpha\Big]\rme^{\frac{1}{2}rp\log p}\Big[ \sum_{\Delta}\!\int\!\rmd t
 \rmd\bx~{W}(\bx|\{\bbeta\})p(t,\Delta|x_0)
 \rme^{- \rmi \sum_{\alpha} \hat{\mathcal{P}}_\alpha(t, \Delta,x_\alpha) }\Big]^n 
 \nonumber
 \\&&\label{eq:inter1}
\end{eqnarray}
in which $\bx=(x_0,\ldots,x_r)\in \mathbb{R}^{r+1}$,  $p(\bz)$ is the distribution from which the $n$ covariate vectors $\bz_i$ were drawn, $p(t,\Delta|\bbeta_0\!\cdot\!\bz)$ is defined in (\ref{eq:data_generation}), and
\begin{eqnarray}
{W}(\bx|\{\bbeta\})&=& \int\!\rmd\bz~p(\bz)\prod_{\alpha=0}^r\delta\Big[x_\alpha-\bbeta_\alpha\!\cdot\!\bz\Big]
\label{eq:score_stats}
\end{eqnarray}
 The integrand in the last line of (\ref{eq:inter1}) depends on the vectors $\bbeta_{\alpha}$ only via the linear predictors $x_\alpha=\bbeta_\alpha\!\cdot\!\bz$. We assume that the distribution $W(\bx|\{\bbeta\})$ of these predictors is Gaussian, either because $p(\bz)$ is Gaussian, or because 
 for $p\to\infty$ the central limit theorem will apply.  Since $\int\!\rmd\bz~p(\bz)\bz={\bf 0}$, 
we may now write, with $\bC=\{C_{\alpha\rho}\}$:
\begin{eqnarray}
&&
\hspace*{-10mm}
W(\bx|\{\bbeta\})=\big[(2\pi)^{r+1}{\rm det}\bC[\{\bbeta\}]\big]^{-\frac{1}{2}}\rme^{-\frac{1}{2}\bx\cdot\bC^{-1}[\{\bbeta\}]\bx}
\\
&&
\hspace*{-10mm}
\forall\alpha,\rho=0\ldots r\!:~~~~ C_{\alpha\rho}[\{\bbeta\}]= \int\!\rmd\bz~p(\bz)(\bbeta_\alpha\!\cdot\!\bz)(\bbeta_\rho\!\cdot\!\bz)=\bbeta_\alpha\!\cdot\bA\bbeta_\rho
\end{eqnarray}
To describe the generation or explanation of finite event times, the linear predictors  will always have to remain finite. Hence for realistic data sets the matrix elements $C_{\alpha\rho}$ must remain of order $O(1)$ even in the limit $p\to\infty$\footnote{In earlier papers \cite{coolen_2017,sheikh,GLM,massa} the definition (\ref{eq:score_stats}) involved $\bbeta_\alpha\!\cdot\!\bz/\sqrt{p}$ instead of $\bbeta_\alpha\!\cdot\!\bz$, so the components of $\bbeta_\alpha$ would typically be of order $\order(1)$. The present choice links more directly  to the standard ML algorithms, but is the reason why in (\ref{eq:intermediate}) the regularizer constant $\frac{1}{2}p\log p$ was required.}. If we now also insert the integral $1=\int\!\rmd\bC~\delta[\bC-\bC[\{\bbeta\}]]$, and write the $\delta$-function in integral form, we find:
\begin{eqnarray}
\hspace*{-15mm}  \big\bra[Z_n(\gamma,\data)]^r\big\ket_{\!\data}&=&\!
 \int\!\rmd\bC\rmd\hat{\bC}\Big[\prod_{\alpha=1}^r\! \mathcal{D}\hat{\mathcal{P}}_\alpha \mathcal{D}\mathcal{P}_\alpha\Big]
 \rme^{-nr\Psi\big[\{{\mathcal P}_\alpha,\hat{\mathcal P}_\alpha\},\bC,\hat{\bC}\big]+\order(\log n)}
  \label{eq:inter2}
 \end{eqnarray}
 where, using $p=\zeta n$, 
 \begin{eqnarray}
 \hspace*{-20mm}
-r \Psi\big[\{{\mathcal P}_\alpha,\hat{\mathcal P}_\alpha\},\bC,\hat{\bC}]&=& 
\rmi \sum_{\alpha=1}^r\sum_{\Delta} \int\rmd t\rmd h~ \hat{\mathcal{P}}_\alpha(t,\Delta,h)\mathcal{P}_\alpha(t,\Delta,h) - 
 \gamma \sum_{\alpha=1}^r\mathcal{E}[\mathcal{P}_\alpha (.)]  \nonumber
 \\
 \hspace*{-20mm}
 &&\hspace*{0mm} +\rmi\zeta {\rm Tr}(\hat{\bC}\bC)+ \varphi[\bC,\{\hat{\mathcal P}_\alpha\}]+\zeta\phi[\hat{\bC}]
 \label{eq:Psi}
\end{eqnarray}
with 
\begin{eqnarray}
\hspace*{-15mm}
\varphi[\bC,\{\hat{\mathcal P}_\alpha\}]&=& \log \Bigg[\sum_{\Delta}\int\!\rmd t\rmd\bx~p(t,\Delta|x_0)\frac{\rme^{-\frac{1}{2}\bx\cdot\bC^{-1}\bx-\rmi\sum_\alpha\hat{\mathcal P}_\alpha(t,\Delta,x_\alpha)}}{\big[(2\pi)^{r+1}{\rm det}\bC\big]^{\frac{1}{2}}}\Bigg]
\\
\hspace*{-15mm}
\phi[\hat{\bC}]&=& \frac{1}{p}\log \int\!\Big[\prod_{\alpha=1}^r \rmd\bbeta_\alpha\Big]~\rme^{\frac{1}{2}rp\log p-\rmi p\sum_{\alpha\rho=0}^r \hat{C}_{\alpha\rho}\bbeta_\alpha\cdot\bA\bbeta_\rho}
\end{eqnarray}
It now follows, after exchanging the limits $n\to\infty$ and $r\to 0$,  that the asymptotic disorder-average free energy density   (\ref{eq:FED}) can be computed via steepest descent:
\begin{eqnarray}
\hspace*{-0mm}
   f(\gamma) &=&\frac{1}{\gamma}\lim_{r\to 0} {\rm extr}_{\{{\mathcal P}_\alpha,\hat{\mathcal P}_\alpha\},\bC,\hat{\bC}}~\Psi\big[\{{\mathcal P}_\alpha,\hat{\mathcal P}_\alpha\},\bC,\hat{\bC}].
   \label{eq:SP_RSB}
   \end{eqnarray}

\subsection{Saddle point integration}

We first derive the stationary conditions for (\ref{eq:Psi}) with respect to the functions $\hat{\mathcal{P}}_{\alpha}$ and ${\mathcal{P}}_{\alpha}$, via functional differentiation and using (\ref{eq:E2}). This gives, respectively, 
\begin{eqnarray}
\hspace*{-15mm}
    \mathcal{P}_{\alpha}(t,\Delta,h)  &=& \frac{\int\!\rmd\bx~\delta[h-x_\alpha]p(t,\Delta|x_0)\rme^{-\frac{1}{2}\bx\cdot\bC^{-1}\bx-\rmi\sum_{\rho=1}^r \hat{\mathcal P}_\rho(t,\Delta,x_\rho)}}{\sum_{\Delta^\prime}\!\int\!\rmd t^\prime\rmd\bx~p(t^\prime\!,\Delta^\prime|x_0)\rme^{-\frac{1}{2}\bx\cdot\bC^{-1}\bx-\rmi\sum_{\rho=1}^r\hat{\mathcal P}_\rho(t^\prime,\Delta^\prime,x_\rho)}}
    \label{eq:RSB_P}
     \\
 \hspace*{-15mm}   \rmi\hat{\mathcal{P}}_{\alpha}(t,\Delta,h)  &=& \gamma  \Delta\Bigg[ \log \Bigg(\sum_{\Delta^\prime}\int\!\rmd t^\prime\rmd h^\prime~\theta(t^\prime-t) \rme^{h'} \mathcal{P}_{\alpha}(t^\prime,\Delta^\prime,h^\prime) \Bigg) - h \Bigg]   \nonumber \\
\hspace*{-10mm}    &&\hspace*{-7mm} + \gamma \rme^{h} \!\int\!\rmd t^\prime\rmd h^\prime \Bigg( \frac{ \theta(t-t^\prime) {\mathcal P}_\alpha(t^\prime,1,h^\prime)}{\sum_{\Delta^\pprime}\int\!\rmd t^\pprime\rmd h^\pprime~ \theta(t^\pprime-t^\prime)\rme^{h^\pprime}\mathcal{P}_{\alpha}(t^\pprime,\Delta^\pprime,h^\pprime) } \Bigg).~~~
\label{eq:RSB_Phat}
\end{eqnarray}
Combination of (\ref{breslow}) with (\ref{eq:Pdensity}) enables us to recognize in (\ref{eq:RSB_Phat}) the replicated cumulative hazard function, 
\begin{eqnarray}
\Lambda_\alpha(t)&=& \int\!\rmd t^\prime\rmd h^\prime~\frac{\theta(t-t^\prime){\mathcal P}_\alpha(t^\prime,1,h^\prime)}{
{\mathcal S}_\alpha(t^\prime)}
\label{eq:RSB_Lambda}
\\
{\mathcal S}_\alpha(t)&=& \sum_{\Delta^\prime}\int\!\rmd t^\prime \rmd h^\prime~\theta(t^\prime-t)\rme^{h^\prime}{\mathcal P}_\alpha(t^\prime,\Delta^\prime,h^\prime)
\label{eq:RSB_S}
\end{eqnarray}
with which we may write (\ref{eq:RSB_Phat}) more compactly as
\begin{eqnarray}
 \rmi\hat{\mathcal{P}}_{\alpha}(t,\Delta,h)  &=& \gamma \Big( \Delta[ \log {\mathcal S}_\alpha(t) - h]  +  \rme^{h} \Lambda_\alpha(t)\Big).
 \label{eq:Phat_compact}
\end{eqnarray}
Insertion into (\ref{eq:RSB_P}) then simplifies the latter to
\begin{eqnarray}
\hspace*{-25mm}
    \mathcal{P}_{\alpha}(t,\Delta,h)  &=& \frac{\int\!\rmd\bx~\delta[h\!-\!x_\alpha]p(t,\Delta|x_0)\rme^{-\frac{1}{2}\bx\cdot\bC^{-1}\bx-\gamma\sum_{\rho=1}^r \!\big( \Delta[ \log {\mathcal S}_\rho(t) - x_\rho]  +  \exp(x_\rho) \Lambda_\rho(t)\big)}}{\sum_{\Delta^\prime}\!\int\!\rmd t^\prime\rmd\bx~p(t^\prime\!,\Delta^\prime|x_0)\rme^{-\frac{1}{2}\bx\cdot\bC^{-1}\bx-\gamma\sum_{\rho=1}^r \!\big( \Delta^\prime[ \log {\mathcal S}_\rho(t^\prime) - x_\rho]  +  \exp(x_\rho) \Lambda_\rho(t^\prime)\big)}}
    \hspace*{-2mm}
    \nonumber
    \\\hspace*{-25mm}&& 
    \label{eq:RSB_P_compact}
     \end{eqnarray}
     Upon using also a modest amount of foresight we transform $\rmi\hat{\bC}=\frac{1}{2}\bD$. We then find that 
our    saddle point problem (\ref{eq:SP_RSB}) can be written as follows:
\begin{eqnarray}
   f(\gamma) &=&\frac{1}{\gamma} \lim_{r\to 0}{\rm extr}_{\bC,\bD,\{{\mathcal P}_{\alpha}\}}~\tilde{\Psi}[\bC,\bD,\{{\mathcal P}_\alpha\}], 
   \label{eq:SP_RSB2}
   \end{eqnarray}     
 where 
    \begin{eqnarray}
&& \hspace*{-25mm}
-r\tilde{\Psi}[\ldots]=  \frac{1}{2}\zeta {\rm Tr}(\bD\bC)+ \tilde{\varphi}[\bC]+\zeta\tilde{\phi}[{\bD}]
-\frac{1}{2}\log{\rm det}\bC-\frac{1}{2}(r\!+\!1)\log(2\pi)
\nonumber
\\
 \hspace*{-20mm}
 &&\hspace*{-10mm} 
+\gamma \sum_{\alpha=1}^r\Big\{
\sum_{\Delta} \int\rmd t\rmd h~ {\mathcal{P}}_\alpha(t,\Delta,h)  \Big( \Delta[ \log {\mathcal S}_\alpha(t) - h]  +  \rme^{h} \Lambda_\alpha(t)\Big) - 
\mathcal{E}[\mathcal{P}_\alpha (.)] \Big\}
\nonumber
\hspace*{-20mm}\\[-1mm]&&
 \label{eq:Psi_RSB2}\\[-10mm]
 \hspace*{-20mm}&&\nonumber
\end{eqnarray}
with 
\begin{eqnarray}
\hspace*{-23.5mm}
\tilde{\phi}[{\bD}]&=& \lim_{p\to\infty}\frac{1}{p}\log \int\!\Big[\prod_{\alpha=1}^r \rmd\bbeta_\alpha\Big]~\rme^{\frac{1}{2}rp\log p-\frac{1}{2} p\sum_{\alpha\rho=0}^r D_{\alpha\rho}\bbeta_\alpha\cdot\bA\bbeta_\rho}
\label{eq:phi}
\\
\hspace*{-23.5mm}
\tilde{\varphi}[\bC]&=& \log \Bigg[\sum_{\Delta}\int\!\rmd t\rmd\bx~p(t,\Delta|x_0)\rme^{-\frac{1}{2}\bx\cdot\bC^{-1}\bx-\gamma\sum_{\alpha=1}^r \big( \Delta[ \log {\mathcal S}_\alpha(t) - x_\alpha]  +  \exp(x_\alpha) \Lambda_\alpha(t)\big)}\Bigg]
\nonumber\\
\hspace*{-25mm}&&\label{eq:varphi}
\end{eqnarray}
We have not used  (\ref{eq:RSB_P_compact}) to derive  (\ref{eq:Psi_RSB2}), only (\ref{eq:Phat_compact}). Hence in   (\ref{eq:SP_RSB2}) the functions $\{{\mathcal P}_\alpha\}$ indeed still have the status of independent variational parameters, and 
 functional differentiation of (\ref{eq:Psi_RSB2}) with respect to ${\mathcal P}_\alpha$ must therefore reproduce (\ref{eq:RSB_P_compact}).

\subsection{The replica symmetry (RS) ansatz}

We now make the so-called replica symmetric (RS) ansatz. Replica symmetry should affect only the nonzero replica labels $\alpha=1\ldots r$, so for the present order parameters $\bC$ and $\bD$ the RS ansatz takes the following form: 
\begin{eqnarray}
\hspace*{-23mm}
\forall \alpha\in\{1,\ldots,r\}\!: &~~& C_{\alpha 0}=C_{0\alpha}=c_0,~~~C_{\alpha\nu}=C_{\nu\alpha}=c+(C-c)\delta_{\alpha\nu}
\\
\hspace*{-20mm}
&& D_{\alpha 0}=D_{0\alpha}=\hat{m},~~D_{\alpha\nu}=D_{\nu\alpha}=-\hat{q}+(\hat{\rho}+\hat{q})\delta_{\alpha\nu}
\\
\hspace*{-20mm}
&&
   {\mathcal P}_\alpha(t,\Delta,h)= {\mathcal P}(t,\Delta,h),~~~ \Lambda_{\alpha}(t)  = \Lambda(t),~~~ {\mathcal S}_\alpha(t)={\mathcal S}(t)
   ~~
\end{eqnarray}
The RS form of $\bC$ implies the same for its inverse, where we define $(\bC^{-1})_{00}=\tilde{\mu}$ and 
\begin{eqnarray}
\hspace*{-20mm}
\forall \alpha\in\{1,\ldots,r\}\!: &~~& (\bC^{-1})_{\alpha 0}=(\bC^{-1})_{0\alpha}=\tilde{m},\\
&& (\bC^{-1})_{\alpha\nu}=(\bC^{-1})_{\nu\alpha}=\tilde{q}+(\tilde{\rho}-\tilde{q})\delta_{\alpha\nu}
\end{eqnarray}
After some algebra one then finds that
\begin{eqnarray}
    \tilde{\mu} &=& \frac{C+c(r-1)}{C_{00}(C +c(r-1))-rc_0^2} \\
    \tilde{m} &=& \frac{c_0}{rc_0^2-C_{00}(C + c(r-1))}\\ 
    \tilde{q} &=& \frac{1}{C-c} \; \frac{c_0^2-cC_{00}}{C_{00}(C+ c(r-1))-rc_0^2} \\
    \tilde{\rho} &=& \frac{1}{C-c}+\tilde{q}~.
\end{eqnarray}
With the RS ansatz we can simplify the functions (\ref{eq:phi},\ref{eq:varphi}). 
We start with $\tilde{\phi}[\bD]$, using the short-hand $S^2=C_{00}=\bbeta_0\cdot\bA\bbeta_0$:
\begin{eqnarray}
\hspace*{-23mm}
\tilde{\phi}_{\rm RS}[{\bD}]&=&\lim_{p\to\infty}\Bigg\{\frac{1}{2}rp\log p-\frac{1}{2}D_{00}\bbeta_0\cdot\bA\bbeta_0
-\frac{r}{2p}\log {\rm Det}\bA
\nonumber
\\
\hspace*{-23mm}
&&
\hspace*{-5mm}
+
 \frac{1}{p}\log \int\!\Big[\prod_{\alpha=1}^r \rmd\bbeta_\alpha\Big]~\rme^{
 -p\hat{m}\sum_{\alpha=1}^r\bbeta_0\cdot\bA^{\frac{1}{2}}\bbeta_\alpha 
+ \frac{1}{2} p\hat{q}[\sum_{\alpha=1}^r \bbeta_\alpha]^2
- \frac{1}{2} p(\hat{\rho}+\hat{q})\sum_{\alpha=1}^r (\bbeta_\alpha)^2}\Bigg\}
\nonumber
\\
\hspace*{-23mm} &=&\lim_{p\to\infty}\Bigg\{\frac{1}{2}rp\log p-\frac{1}{2}D_{00}\bbeta_0\cdot\bA\bbeta_0
-\frac{r}{2p}\log {\rm Det}\bA
\nonumber
\\
\hspace*{-23mm}
&&
+
 \frac{1}{p}\log \int\!\Big[\prod_{\mu=1}^p {
 \rm D}x_\mu\Big]\Big[\prod_{\mu=1}^p \int\!\rmd\beta~\rme^{
  \beta[x_\mu\sqrt{p\hat{q}} -p\hat{m}(\bA^{\frac{1}{2}}\bbeta_0)_\mu]
- \frac{1}{2} p(\hat{\rho}+\hat{q}) \beta^2}
\Big]^r\Bigg\}
\nonumber
\\
\hspace*{-23mm} &=&-\frac{1}{2}D_{00}S^2+
\frac{1}{2}r\Big(\frac{\hat{m}^2 S^2\!+\!\hat{q}}{\hat{q}\!+\!\hat{\rho}}\!-\frac{1}{p}\log {\rm Det}\bA\!-\log(\hat{\rho}\!+\!\hat{q})-\log(2\pi)\Big)+\order(r^2)
\nonumber
\\[-1mm]
\hspace*{-23mm}
\end{eqnarray}
Next we apply the RS ansatz to $\tilde{\varphi}[\bC]$:
\begin{eqnarray}
\hspace*{-20mm}
\rme^{\tilde{\varphi}_{\rm RS}[\bC]}&=& \sum_{\Delta\in\{0,1\}}
\int\!\rmd t\rmd\bx~p(t,\Delta|x_0)\rme^{
 \Delta \gamma\sum_{\alpha=1}^rx_\alpha
-\gamma \Lambda(t)\sum_{\alpha=1}^r \exp(x_\alpha)-r\gamma \Delta \log {\mathcal S}(t)}
\nonumber
\\[-1mm]
\hspace*{-20mm}
&&\hspace*{20mm}\times \rme^{-\frac{1}{2}\tilde{\mu}x_0^2-\tilde{m}x_0\sum_{\alpha=1}^rx_\alpha-\frac{1}{2}\tilde{q}[\sum_{\alpha=1}^r x_\alpha]^2-\frac{1}{2}(\tilde{\rho}-\tilde{q})\sum_{\alpha=1}^r x_\alpha^2}
\nonumber
\\[1mm]
\hspace*{-20mm}
&=& \sum_{\Delta\in\{0,1\}}
\int\!{\rm D}z\!\int\!\rmd t \rmd x_0~\rme^{-\frac{1}{2}\tilde{\mu}x_0^2} p(t,\Delta|x_0)\rme^{-r\gamma \Delta \log {\mathcal S}(t)}
\nonumber
\\[-1mm]
\hspace*{-20mm}&&
\hspace*{20mm} \times
\Big[\int\!\rmd x~
 \rme^{x(\Delta \gamma +\rmi z\sqrt{\tilde{q}}-\tilde{m}x_0)
-\gamma \Lambda(t) \exp(x)-\frac{1}{2}(\tilde{\rho}-\tilde{q})x^2}\Big]^r
\nonumber
\\
\hspace*{-20mm}&=&
\Big(\frac{2\pi}{\tilde{\mu}}\Big)^{\!\frac{1}{2}}\Bigg\{1
-r\gamma\int\!{\rm D}y_0 \int\!\rmd t~ p(t,1|\frac{y_0}{\sqrt{\tilde{\mu}}}) \log {\mathcal S}(t)
+\frac{1}{2}r\log \Big(\frac{2\pi}{\tilde{\rho}-\tilde{q}}\Big)+\order(r^2)
\nonumber
\\
\hspace*{-20mm}&&
+r\!\sum_{\Delta\in\{0,1\}}
\int\!{\rm D}y_0{\rm D}z\!\int\!\rmd t~ p(t,\Delta|\frac{y_0}{\sqrt{\tilde{\mu}}})
\nonumber
\\[-1mm]
\hspace*{-20mm}&&
\hspace*{5mm} 
\times \log\int\!\rmD x~
 \rme^{x[\Delta \gamma +\rmi z\sqrt{\tilde{q}}-(\tilde{m}/\sqrt{\tilde{\mu}})y_0]/\sqrt{\tilde{\rho}-\tilde{q}}
-\gamma \Lambda(t) \exp(x/\sqrt{\tilde{\rho}-\tilde{q}})}
\end{eqnarray}
Hence
\begin{eqnarray}
\hspace*{-20mm}
\tilde{\varphi}_{\rm RS}[\bC]&=&\frac{1}{2}\log 
\Big(\frac{2\pi}{\tilde{\mu}}\Big)
-r\gamma\int\!{\rm D}y_0  \int\!\rmd t~ p(t,1|\frac{y_0}{\sqrt{\tilde{\mu}}}) \log {\mathcal S}(t)
+\frac{1}{2}r\log \Big(\frac{2\pi}{\tilde{\rho}-\tilde{q}}\Big)+\order(r^2)
\nonumber
\\
\hspace*{-20mm}&&
+r\!\sum_{\Delta\in\{0,1\}}
\int\!{\rm D}y_0{\rm D}z\!\int\!\rmd t~ p(t,\Delta|\frac{y_0}{\sqrt{\tilde{\mu}}})
\nonumber
\\[-1mm]
\hspace*{-20mm}&&
\hspace*{5mm} 
\times \log\int\!\rmD x~
 \rme^{x[\Delta \gamma +\rmi z\sqrt{\tilde{q}}-(\tilde{m}/\sqrt{\tilde{\mu}})y_0]/\sqrt{\tilde{\rho}-\tilde{q}}
-\gamma \Lambda(t) \exp(x/\sqrt{\tilde{\rho}-\tilde{q}})}
\end{eqnarray}
We then find, using identities such as ${\rm Tr}(\bD\bC)=D_{00}S^2+r(2c_0\hat{m}\!+\!c\hat{q}\!+\!C\hat{\rho})+\order(r^2)$, $\log{\rm Det}\bC=\log S^2+r[\log(C\!-\!c)+(c\!-\!c_0^2/S^2)/(C\!-\!c)]+\order(r^2)$ (obtained by diagonalizing the RS form of $\bC$), as well as $\tilde{\mu}S^2= 1+rc_0^2/S^2(C\!-\!c)+\order(r^2)$, $\tilde{m}=-c_0/S^2(C\!-\!c)+\order(r)$, $\tilde{q}=(c_0^2/S^2\!-\!c)/(C\!-\!c)^2+\order(r)$, and $\tilde{\rho}-\tilde{q}=(C\!-\!c)^{-1}+\order(r)$,  that the RS ansatz implies the following for expression (\ref{eq:Psi_RSB2}) (modulo irrelevant constants):
  \begin{eqnarray}
&& \hspace*{-26mm}
-\lim_{r\to 0}\tilde{\Psi}_{\rm RS}[\ldots]= 
-\frac{1}{2}\frac{c}{C\!-\!c}
+\frac{1}{2}\zeta \Big(
2\hat{m}c_0\!+\!\hat{q}c\!+\!\hat{\rho}C+
\frac{\hat{m}^2 S^2\!+\!\hat{q}}{\hat{q}\!+\!\hat{\rho}}\!-\log(\hat{\rho}\!+\!\hat{q})\Big)
\nonumber
\\
\hspace*{-20mm}&&
\hspace*{-25mm}
+\sum_{\Delta}
\int\!{\rm D}y_0{\rm D}z\!\int\!\!\rmd t~ p(t,\Delta|Sy_0)
 \log\!\int\!\!\rmD x~
 \rme^{\big[\Delta \gamma \sqrt{C\!-c}+z\frac{\sqrt{\!c-c_0^2/\!S^2}}{\sqrt{C\!-c}}+\frac{y_0c_0}{S\sqrt{C\!-c}}\big]x
-\gamma \Lambda(t) \exp(x\sqrt{C-c})}
\nonumber\black
\\
 \hspace*{-20mm}
 &&\hspace*{-0mm} 
+\gamma \Bigg\{
\sum_{\Delta\in\{0,1\}} \int\rmd t\rmd h~ {\mathcal{P}}(t,\Delta,h)  \Big( \Delta[ \log {\mathcal S}(t) - h]  +  \rme^{h} \Lambda(t)\Big) - 
\mathcal{E}[\mathcal{P}(.)] 
\nonumber
\\[-1mm]
&&\hspace*{35mm}-\int\!{\rm D}y_0  \int\!\rmd t~ p(t,1|Sy_0) \log {\mathcal S}(t)\Bigg\}~.
\end{eqnarray}
We can now extremize $\Psi_{\rm RS}[\ldots]$ over $(\hat{m},\hat{q},\hat{\rho})$, as demanded by (\ref{eq:SP_RSB2}), giving
\begin{eqnarray}
\hat{m}=-\frac{c_0}{S^2(C\!-\!c)},~~~~~\hat{q}=\frac{c\!-\!c_0^2/S^2}{(C\!-\!c)^2},~~~~~\hat{\rho}\!+\!\hat{q}=\frac{1}{C\!-\!c}
\end{eqnarray}
Our various formulae takes more compact forms upon introducing the three short-hands $u=\sqrt{\gamma(C\!-\!c)}$, $v=\sqrt{c\!-\!c_0^2/S^2}$ and $w=c_0/S$, which we know from earlier studies to have finite $\gamma\to\infty$ limits. 
We may then write
\begin{eqnarray}
\hat{m}=-\frac{\gamma w}{Su^2},~~~~~\hat{q}=\frac{\gamma^2 v^2}{u^4},~~~~~\hat{\rho}\!+\!\hat{q}=\frac{\gamma}{u^2}
\end{eqnarray}
and
\begin{eqnarray}
&& \hspace*{-26mm}
-\lim_{\gamma\to\infty}\frac{1}{\gamma}\lim_{r\to 0}\tilde{\Psi}_{\rm RS}[\ldots]= 
-\frac{1}{2u^2}[w^2\!+\!(1\!-\!\zeta)v^2]
\nonumber
\\
\hspace*{-20mm}&&
\hspace*{-23mm}+\sum_{\Delta}\!
\int\!{\rm D}y_0{\rm D}z\rmd t~ p(t,\Delta|Sy_0)
\lim_{\gamma\to\infty}\!\frac{1}{\gamma} \log\!\int\!\rmd x~
 \rme^{\gamma\Big[-\frac{1}{2}u^2x^2+[\Delta u^2+vz+wy_0]x
- \Lambda(t) \exp(xu^2)\Big]}
\nonumber\black
\\
 \hspace*{-20mm}
 &&\hspace*{-0mm} 
+
\sum_{\Delta\in\{0,1\}} \int\rmd t\rmd h~ {\mathcal{P}}(t,\Delta,h)  \Big( \Delta[ \log {\mathcal S}(t) - h]  +  \rme^{h} \Lambda(t)\Big) - 
\mathcal{E}[\mathcal{P}(.)] 
\nonumber
\\[-1mm]
&&\hspace*{35mm}-\int\!{\rm D}y_0  \int\!\rmd t~ p(t,1|Sy_0) \log {\mathcal S}(t)~.
\end{eqnarray}
We may hence now write $f_{\rm RS}=\lim_{\gamma\to\infty} f_{\rm RS}(\gamma)$ as follows:
\begin{eqnarray}
\hspace*{-25mm}
f_{\rm RS}&=&{\rm extr}_{u,v,w,{\mathcal P}}\Bigg\{
\frac{1}{2u^2}[w^2\!+\!(1\!-\!\zeta)v^2]
\nonumber
\\
\hspace*{-25mm}&&
\hspace*{-3mm}-\!\!\sum_{\Delta\in\{0,1\}}\!
\int\!{\rm D}y{\rm D}z\rmd t~ p(t,\Delta|Sy)~{\rm max}_{x}
\Big[\!-\!\frac{1}{2}u^2x^2+(\Delta u^2\!+\!vz\!+\!wy)x
- \Lambda(t) \rme^{xu^2}\Big]
\nonumber\black
\\
 \hspace*{-25mm}
 &&\hspace*{22mm} 
-\!\!
\sum_{\Delta\in\{0,1\}} \int\rmd t\rmd h~ {\mathcal{P}}(t,\Delta,h)  \Big( \Delta[ \log {\mathcal S}(t) - h]  +  \rme^{h} \Lambda(t)\Big) - 
\mathcal{E}[\mathcal{P}(.)] 
\nonumber
\\[-1mm]
\hspace*{-20mm}
&&\hspace*{45mm}+\int\!{\rm D}y  \int\!\rmd t~ p(t,1|Sy) \log {\mathcal S}(t)
\Bigg\}
\label{eq:fRS_nearly}
\end{eqnarray}
with  $\mathcal{E}[\mathcal{P}(.)]$ as defined by  (\ref{eq:E2}), and 
\begin{eqnarray}
\Lambda(t)&=& \int\!\rmd t^\prime\rmd h^\prime~\frac{\theta(t-t^\prime){\mathcal P}(t^\prime,1,h^\prime)}{
{\mathcal S}(t^\prime)}
\label{eq:RS_Lambda}
\\
{\mathcal S}(t)&=& \sum_{\Delta^\prime}\int\!\rmd t^\prime \rmd h^\prime~\theta(t^\prime-t)\rme^{h^\prime}{\mathcal P}(t^\prime,\Delta^\prime,h^\prime)
\label{eq:RS_S}
\end{eqnarray}
The maximization over $x$ in (\ref{eq:fRS_nearly}) is achieved for $x=u^{-2}\xi(t,\Delta,y,z)$, where
\begin{eqnarray}
\xi(t,\Delta,y,z)=\Delta u^2\!+\!vz\!+\!wy-W\Big(u^2\Lambda(t)\rme^{\Delta u^2+vz+wy}\Big)
\label{eq:define_xi}
\end{eqnarray}
with Lambert's function $W(z)$, the inverse of the function $g(x)=x\rme^x$. 
Note that expression (\ref{eq:define_xi}) can be interpreted in terms of a Moreau envelope, and is in other approaches to overfitting  indeed often found via that route \cite{loureiro}. 
Hence we now arrive at
\begin{eqnarray}
\hspace*{-25mm}
f_{\rm RS}&=&{\rm extr}_{u,v,w,{\mathcal P}}\Bigg\{
\frac{1}{2u^2}[w^2\!+\!(1\!-\!\zeta)v^2]
+\frac{1}{u^2}\!\!\sum_{\Delta\in\{0,1\}}\!
\int\!{\rm D}y{\rm D}z\rmd t~ p(t,\Delta|Sy)
\Bigg[\frac{1}{2}\xi^2(t,\Delta,y,z)
\nonumber
\\
\hspace*{-25mm}&&\hspace*{45mm} -(\Delta u^2\!+\!vz\!+\!wy)\xi(t,\Delta,y,z)
+u^2\Lambda(t) \rme^{\xi(t,\Delta,y,z)}\Bigg] 
\nonumber
\\
\hspace*{-25mm}
&&
-\!\!
\sum_{\Delta\in\{0,1\}} \int\rmd t\rmd h~ {\mathcal{P}}(t,\Delta,h)  \Big( \Delta[ \log {\mathcal S}(t) - h]  +  \rme^{h} \Lambda(t)\Big) - 
\mathcal{E}[\mathcal{P}(.)] 
\nonumber
\\[-1mm]
\hspace*{-20mm}
&&\hspace*{45mm}+\int\!{\rm D}y  \int\!\rmd t~ p(t,1|Sy) \log {\mathcal S}(t)
\Bigg\},
\label{eq:fRS}
\end{eqnarray}
which can be written in alternative ways, using identities such as $z\rme^{-W(z)}=W(z)$.

\subsection{Replica symmetric order parameter equations}

We next work out the RS form of equation (\ref{eq:RSB_P_compact}) in the limit $r\to 0$, using the same manipulations as in working out the previous function $\tilde{\varphi}_{\rm RS}[\bC]$, giving
\begin{eqnarray}
\hspace*{-25mm}
    \mathcal{P}(t,\Delta,h)  &=&\lim_{r\to 0} \Bigg[
   \int\!{\rm D}y{\rm D}z~p(t,\Delta|\frac{y}{\sqrt{\tilde{\mu}}})\rme^{-r\gamma\Delta \log {\mathcal S}(t)}~\frac{\sqrt{\tilde{\rho}-\tilde{q}}}{\sqrt{2\pi}} \rme^{-\frac{1}{2}(\tilde{\rho}-\tilde{q})h^2}  \nonumber
   \\
   \hspace*{-25mm}&& \hspace*{-10mm}\times
\rme^{[\Delta\gamma+\rmi z\sqrt{\tilde{q}} -\frac{\tilde{m}y}{\sqrt{\tilde{\mu}}}] h -\gamma \Lambda(t)\exp(h) }    
    \Big(\int\!{\rm D}x~
    \rme^{[\Delta\gamma+\rmi z\sqrt{\tilde{q}} -\frac{\tilde{m}y}{\sqrt{\tilde{\mu}}}] \frac{x}{\sqrt{\tilde{\rho}-\tilde{q}}} -\gamma \Lambda(t)\exp( \frac{x}{\sqrt{\tilde{\rho}-\tilde{q}}}) }\Big)^{r-1}
    \Bigg]
    \nonumber
    \\
    \hspace*{-25mm}
    &&\times\Bigg[
    \sum_{\Delta^\prime\in\{0,1\}}\int\!\rmd t^\prime
    \int\!{\rm D}y{\rm D}z~p(t^\prime,\Delta^\prime|\frac{y}{\sqrt{\tilde{\mu}}})\rme^{-r\gamma\Delta^\prime \log {\mathcal S}(t^\prime)}
    \nonumber
    \\[-1mm]
    \hspace*{-25mm}&&\hspace*{30mm}\times
    \Big(\int\!{\rm D}x~
    \rme^{[\Delta^\prime\gamma+\rmi z\sqrt{\tilde{q}} -\frac{\tilde{m}y}{\sqrt{\tilde{\mu}}}] \frac{x}{\sqrt{\tilde{\rho}-\tilde{q}}} -\gamma \Lambda(t^\prime)\exp( \frac{x}{\sqrt{\tilde{\rho}-\tilde{q}}}) }\Big)^{r}
    \Bigg]^{-1}
    \nonumber
    \\
    \hspace*{-25mm}
    &=& \lim_{r\to 0}
     \int\!{\rm D}y{\rm D}z~p(t,\Delta|\frac{y}{\sqrt{\tilde{\mu}}})\frac{\frac{\sqrt{\tilde{\rho}-\tilde{q}}}{\sqrt{2\pi}} \rme^{-\frac{1}{2}(\tilde{\rho}-\tilde{q})h^2+[\Delta\gamma+\rmi z\sqrt{\tilde{q}} -\frac{\tilde{m}y}{\sqrt{\tilde{\mu}}}] h -\gamma \Lambda(t)\exp(h) }    }
    {\int\!{\rm D}x~
    \rme^{[\Delta\gamma+\rmi z\sqrt{\tilde{q}} -\frac{\tilde{m}y}{\sqrt{\tilde{\mu}}}] \frac{x}{\sqrt{\tilde{\rho}-\tilde{q}}} -\gamma \Lambda(t)\exp( \frac{x}{\sqrt{\tilde{\rho}-\tilde{q}}}) }}
    \Bigg]
    \nonumber
    \\\
    \hspace*{-25mm} &=&
     \int\!{\rm D}y{\rm D}z~p(t,\Delta|Sy)\frac{\rme^{\frac{\gamma}{u^{2}}\big[-\frac{1}{2}h^2+(\Delta u^2+vz+wy)h-u^2\Lambda(t)\rme^h\big]}}
     {\int\!\rmd h^\prime~\rme^{\frac{\gamma}{u^{2}}\big[-\frac{1}{2}h^{\prime 2}+(\Delta u^2+vz+wy)h^\prime-u^2\Lambda(t)\rme^{h^\prime}\big]}}     
      \end{eqnarray}
      Upon taking the limit $\gamma\to\infty$ we then find the following remarkably simple result, in which the function $\xi(\ldots)$ is as given by equation 
      (\ref{eq:define_xi}): 
            \begin{eqnarray}
 \lim_{\gamma\to\infty}   \mathcal{P}(t,\Delta,h)  &=&      
         \int\!{\rm D}y{\rm D}z~p(t,\Delta|Sy)~\delta\big[h-\xi(t,\Delta,y,z)\big]    .
      \end{eqnarray}
     From this one then obtains the RS expressions of $\Lambda(t)$ and ${\mathcal S}(t)$, via 
     (\ref{eq:RS_Lambda},
\ref{eq:RS_S}). 
\vsp

Finally we need to work out the scalar order parameter equations for the trio $(u,v,w)$, by executing the extremization demanded in expression 
(\ref{eq:fRS}) for the free energy density.  These equations are seen to take the form $\partial{\cal F}(u,v,w)/\partial u=\partial{\cal F}(u,v,w)/\partial v=\partial{\cal F}(u,v,w)/\partial w=0$, with $\xi(\ldots)$ as given in (\ref{eq:define_xi}), where
\begin{eqnarray}
\hspace*{-25mm}
{\cal F}(u,v,w)&=&
\frac{1}{2u^2}[w^2\!+\!(1\!-\!\zeta)v^2]
+\frac{1}{u^2}\!\!\sum_{\Delta\in\{0,1\}}\!
\int\!{\rm D}y{\rm D}z\rmd t~ p(t,\Delta|Sy)
\Big[\frac{1}{2}\xi^2(t,\Delta,y,z)
\nonumber
\\
\hspace*{-25mm}&&\hspace*{25mm} -(\Delta u^2\!+\!vz\!+\!wy)\xi(t,\Delta,y,z)
+u^2\Lambda(t) \rme^{\xi(t,\Delta,y,z)}\Big] 
\label{eq:RS_F_app}
\end{eqnarray}
By using identities such as $W(x)\exp[W(x)]=x$, one finds that $u^2\Lambda(t)\exp[\xi(t,\Delta,y,z)]=W(u^2\Lambda(t)\rme^{\Delta u^2+vz+wy})$, which enables us to simplify ${\cal F}(u,v,w)$ to
\begin{eqnarray}
\hspace*{-25mm}
{\cal F}(u,v,w)&=&\frac{1}{u^2}\Bigg\{
\frac{1}{2}[w^2\!\!+\!(1\!-\!\zeta)v^2]
+\sum_{\Delta}\!
\int\!{\rm D}y{\rm D}z\rmd t~ p(t,\Delta|Sy)
\Big[\frac{1}{2}W^2(u^2\Lambda(t)\rme^{\Delta u^2+vz+wy})\nonumber
\\
\hspace*{-25mm}&&\hspace*{25mm} +W(u^2\Lambda(t)\rme^{\Delta u^2+vz+wy})-\frac{1}{2}(\Delta u^2\!+\!vz\!+\!wy)^2
\Big] 
\Bigg\}
\end{eqnarray}
From this expression one then derives directly the final order parameter equations for $(u,v,w)$, in which only the equation for $u$ involves some further manipulations: 
\begin{eqnarray}
\hspace*{-15mm}
\frac{\partial{\cal F}}{\partial v}\!=0\!: &~~~~& \zeta v=\int\!{\rm D}y{\rm D}z ~z\sum_\Delta\!\int\!\rmd t~ p(t,\Delta|Sy)~W(u^2\Lambda(t)\rme^{\Delta u^2+vz+wy})\nonumber
\\[-1mm]
\hspace*{-15mm}
\label{eq:RS_v_app}
\\[2mm]
\hspace*{-15mm}
\frac{\partial{\cal F}}{\partial w}\!=0\!: && 0~~=\int\!{\rm D}y{\rm D}z~y\sum_\Delta\!\int\!\rmd t~ p(t,\Delta|Sy)~\Big[W(u^2\Lambda(t)\rme^{\Delta u^2+vz+wy})\!-\!\Delta u^2\Big]
\nonumber
\\[-1mm]
\hspace*{-15mm}
\label{eq:RS_w_app}
\\[2mm]
\hspace*{-15mm}
\frac{\partial{\cal F}}{\partial u}\!=0\!: && \zeta v^2\!=\int\!{\rm D}y{\rm D}z\sum_\Delta\!\int\!\rmd t~ p(t,\Delta|Sy)\Big[W(u^2\Lambda(t)\rme^{\Delta u^2+vz+wy})\!-\!\Delta u^2\Big]^2
\nonumber
\\[-1mm]
\hspace*{-15mm}
\label{eq:RS_u_app}
\end{eqnarray}
Finally, after some simple manipulations and usage of (\ref{eq:RS_u_app}) one finds that at the saddle point $(u_\star,v_\star,w_\star)$ the value of ${\cal F}$ in expression (\ref{eq:RS_F_app}) simplifies to
\begin{equation}
\hspace*{-20mm}
    \mathcal{F}(u_\star,v_\star,w_\star)=\int\!{\rm D}y{\rm D}z\! \sum_{\Delta}\int\!\rmd t~p(t,\Delta|Sy)\Big[ \Lambda(t)\rme^{\xi_{\star}(t,\Delta,y,z) } -\Delta\xi_{\star}(t,\Delta,y,z) \Big],~~~
\end{equation}

\section{Variance of inferred risk factors}
\label{app:get_S}

In this Appendix we derive equation (\ref{eq:get_S}) for the variance of the inferred risk factors $\hat{\bbeta}_{\rm ML}\!\cdot\!\bz_i$, which  is an important tool with which to remove the need for knowing the amplitude $S$ in the construction of unbiased estimators. It exploits our result (\ref{eq:RS_P_final}) for the asymptotic form of the distribution (\ref{eq:Pdensity}) (which is the type of quantity that within the replica theory is assumed to be self-averaging\footnote{This assumption is validated by the accuracy of the predictions of the RS equations, as confirmed repeatedly in previous studies \cite{coolen_2017,GLM,massa,massa_jacknife}, and also again in our present numerical simulations.}):
\begin{eqnarray}
\hspace*{-15mm}
\lim_{n\to\infty} \frac{1}{n}\sum_{i=1}^n (\hat{\bbeta}_{\rm ML}\!\cdot\!\bz_i)^2&=& 
\lim_{n\to\infty}\!\sum_{\Delta\in\{0,1\}}\int\!\rmd t\rmd h~\big\bra P_n(t,\Delta,h|\hat{\bbeta}_{\rm ML},\data)\big\ket_{\data} h^2
\nonumber
\\
\hspace*{-15mm}
&=& \sum_{\Delta\in\{0,1\}}\int\!\rmd t\rmd h~{\mathcal P}(t,\Delta,h) h^2
\nonumber
\\
\hspace*{-15mm}
&=& \sum_{\Delta\in\{0,1\}}\int\!{\rm D}y{\rm D}z\rmd t ~p(t,\Delta|Sy)\xi^2(t,\Delta,y,z).
\end{eqnarray}
Upon using (\ref{rs4})  we can thus write 
\begin{eqnarray}
\hspace*{-20mm}
\lim_{n\to\infty} \frac{1}{n}\sum_{i=1}^n (\hat{\bbeta}_{\rm ML}\!\cdot\!\bz_i)^2&=& 
 \sum_{\Delta}\int\!{\rm D}y{\rm D}z\rmd t ~p(t,\Delta|Sy)\nonumber
 \\[-1mm]
 \hspace*{-20mm}&&\hspace*{15mm}\times
 \Big[wy\!+\!vz\!+\!u^2\Delta-W\Big(u^2\rme^{u^2\Delta+wy+vz}\Lambda(t)\Big)\Big]^2
 \nonumber
 \\
 \hspace*{-20mm}&&\hspace*{-25mm}= v^2\!+w^2\!+\! \int\!{\rm D}y{\rm D}z \sum_{\Delta}\int\!\rmd t~p(t,\Delta|Sy)
  \Big[u^2\Delta-W\Big(u^2\rme^{u^2\Delta+wy+vz}\Lambda(t)\Big)\Big]^2\nonumber
  \\
  \hspace*{-20mm}&& \hspace*{-20mm} -2w\int\!{\rm D}y{\rm D}z \sum_{\Delta}\int\!\rmd t~p(t,\Delta|Sy)\Big[yW\Big(u^2\rme^{u^2\Delta+wy+vz}\Lambda(t)\Big)-\Delta u^2 y\Big] 
  \nonumber
  \\
  \hspace*{-20mm} && \hspace*{-20mm} -2v\int\!{\rm D}y{\rm D}z \sum_{\Delta}\int\!\rmd t~p(t,\Delta|Sy)~zW\Big(u^2\rme^{u^2\Delta+wy+vz}\Lambda(t)\Big). 
  \end{eqnarray}
 Upon using the order parameter equations (\ref{eq:RS_v1},\ref{eq:RS_w1},\ref{eq:RS_u1}) we can now simplify the right-hand side of the latter expression, and find the remarkably simple but powerful result
 \begin{eqnarray}
\lim_{n\to\infty} \frac{1}{n}\sum_{i=1}^n (\hat{\bbeta}_{\rm ML}\!\cdot\!\bz_i)^2&=& (\zeta\!-\!1)v_\star^2+w_\star^2.
\label{eq:get_S_app}
\end{eqnarray}
En passant, expression (\ref{eq:get_S_app}) also implies an inequality with which to simplify our numerical solution of the RS equations, namely $w_\star^2\geq (1\!-\!\zeta)v_\star^2$ (note that for ML regression in Generalized Linear Models such as the present, any theory would be applicable only for $\zeta<1$ on dimensional grounds).   

\end{document}